\documentclass[nonblindrev]{informs3}
\usepackage{balance}
\usepackage{amsmath}
\usepackage[tight,footnotesize]{subfigure}
\usepackage{graphicx}
\usepackage{bm}
\usepackage{algorithm}
\usepackage[noend]{algorithmic}
\usepackage{multirow}
\usepackage{diagbox}
\usepackage{caption}
\usepackage{amsfonts}
\usepackage{lettrine}
\OneAndAHalfSpacedXII 

\usepackage{natbib}
 \bibpunct[, ]{(}{)}{,}{a}{}{,}%
 \def\bibfont{\small}%
\TheoremsNumberedThrough     

\EquationsNumberedThrough    

\MANUSCRIPTNO{2015} 

\begin{document}


\RUNAUTHOR{Tang et al.}

\RUNTITLE{Profit-Driven  Team Grouping in Social Networks}

\TITLE{Profit-Driven Team Grouping in Social Networks}

\ARTICLEAUTHORS{%
\AUTHOR{Shaojie Tang}
\AFF{Naveen Jindal School of Management, The University of Texas at Dallas}
\AUTHOR{Jing  Yuan}
\AFF{Department of Computer Science and Engineering, The University of North Texas}
\AUTHOR{Tao Li, Yao Wang}
\AFF{Center of Intelligent Decision-making and Machine Learning,
School of Management, Xi'an Jiaotong University}
} 

\ABSTRACT{In this paper, we investigate the profit-driven team grouping problem in social networks. We consider a setting in which people possess different skills, and the compatibility between these individuals is captured  by a social network.  Moreover, there is a collection of tasks, where each task requires a specific set of skills and yields a profit upon completion. Individuals may collaborate with each other as \emph{teams} to accomplish a set of tasks. We aim  to find a group of teams to maximize the total profit of the tasks that they can complete. Any feasible grouping must satisfy the following conditions: (i) each team possesses all the skills required by the task assigned to it, (ii) individuals belonging to the same team are socially compatible, and (iii) no individual is overloaded. We refer to this as the \textsc{TeamGrouping} problem. We  analyze the computational complexity of this problem and then propose a linear program-based approximation algorithm to address it and its variants. Although we focus on team grouping, our results apply to a broad range of optimization problems that can be formulated as cover decomposition problems.
}


\maketitle

%

\section{Introduction}\label{sec:introduction}
In this paper, we address the  team grouping problem in a networked
community of people with diverse skill sets. We consider a setting where people possess different skills and the compatibility between these individuals is captured  by a social network.  We assume a collection of tasks where each task requires a specific set of skills and yields a profit upon completion. Individuals may collaborate with each other as \emph{teams} to accomplish a set of tasks. We aim to find a grouping method that maximizes the total profit of the tasks they can complete. Relevant examples are available in the domain of online labor markets, such as Freelancer (\verb"www.freelancer.com"), Upwork (\verb"www.upwork.com"), and Guru (\verb"www.guru.com"). In these
online platforms, freelancers with various skills can be hired to work on different types of projects. Instead of working purely independently, a growing number of freelancers are realizing the benefit of working as a team, with fellow freelancers who have complementary skills \citep{golshan2014profit}. This allows them to  expand their talent pool and better balance their workload. Many major platforms in this area, such as Upwork,  provide team hiring services to their enterprise customers.

We formalize the profit-driven team grouping problem as follows:
we assume a set of $m$ individuals $\mathcal{V}$ and a set of
$n$ skills $\mathcal{S}$. Each individual $u \in \mathcal{V}$ is represented by a subset
of skills  possessed by this individual, that is, $u \subseteq S$; these are the skills that the individual
possesses. There is a set of tasks $\mathcal{T}$, and every task $t \in \mathcal{T}$ can also be represented
by the set of skills required by this task (i.e., $t \subseteq \mathcal{S}$). Finally, every task $t$ will yield a profit $\lambda_t$, which is the benefit that the completion of the task will yield for the platform. The team grouping problem (labeled \textsc{TeamGrouping}) is to group individuals into different teams and assign a task to each team in a manner that satisfies  the following conditions: (i) each team possesses all the skills required by the task, (ii) individuals within the same team have high social compatibility, and (iii) no individual is overloaded. Our goal is to maximize the sum of profits  from all the tasks that can be performed by these teams. Social compatibility  between individuals can be interpreted in many ways. In this work, we model \emph{social compatibility}  by means of a social network in which the nodes represent individuals and an edge connecting two
nodes denotes a social connection between  the corresponding individuals. One popular indicator of social compatibility  is \emph{connectivity} \citep{lappas2009finding}; therefore,  each team must form a connected graph. Another important indicator of social compatibility is \emph{diameter}, for example, according to \citep{anagnostopoulos2012online},  the induced graph of each team should have a small diameter. However, our results are not
restricted to any specific indicator of social compatibility. Instead, we propose a
general framework in which a socially compatible team is a subset of nodes of the
graph for which the induced subgraph has some desirable property.

\begin{figure}[tbhp]
\begin{center}
\includegraphics[width=2.5in]{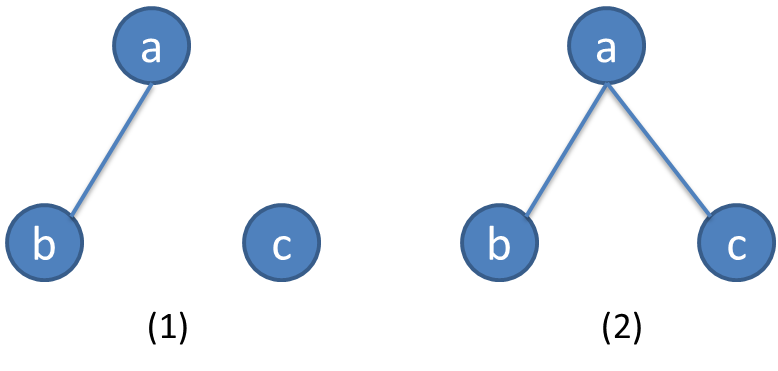}
\caption{Two social networks.}
\label{fig:FSSILQI}
\end{center}
\end{figure}

We next present a toy example of our problem. Assume there are three IT projects requiring different skills: the first task will yield profit $\lambda_1=\$ 50$ and requires skills $t_1 =\{ \mbox{HTML, MySQL, JavaScript, PHP}\}$, the second task will yield profit $\lambda_2=\$ 10$ and requires skills $t_2=\{ \mbox{JavaScript, HTML}\}$, and the last task will yield profit $\lambda_3=\$5$ and requires skills $t_3=\{\mbox{PHP}\}$. In addition, there are three individuals $\{a, b, c\}$  with the following skills: $a= \{\mbox{HTML, MySQL}\}$, $b=\{\mbox{JavaScript}\}$, and $c=\{\mbox{HTML, PHP}\}$.  In our basic formulation, each individual can  participate in only one team, and all team members must be connected. We consider the social networks illustrated in Fig. \ref{fig:FSSILQI}.  The most profitable grouping approach in Fig. \ref{fig:FSSILQI} (1) is to assign team $\{a, b\}$ to $t_2$, and team $\{c\}$ to $t_3$, which yields $\$ 15$ in profit. This is because $a$ and $b$ are connected while $c$ is isolated. For the social network in Fig. \ref{fig:FSSILQI} (2) by contrast, since the induced graph of all three individuals is connected, the most profitable grouping approach is to assign team $\{a, b, c\}$ to $t_1$, which yields $\$ 50$ in profit.

\textbf{Contributions:} To the best of our knowledge, we are the
first to define and study the \textsc{TeamGrouping} problem and its
variants. We summarize our contributions as follows:

\begin{itemize}
\item We show that this problem is $1/\ln m$-hard to approximate; that is, it is NP-hard to find a solution with approximation ratio larger than $1/\ln m$.
\item We propose a linear program (LP) based algorithm with approximation ratio $\max\{\mu/\Delta, \mu  /2\sqrt{m}\}$ where $\Delta$ denotes the size of the largest minimal team and $1/\mu$ is the approximation ratio of the \textsc{MincostTeamSelection} problem (Definition \ref{def:min}). If there is no constraint on social compatibility, then this ratio reduces to $\max\{\ln n/n, \ln n /2\sqrt{m}\}$.
\item We consider two extensions of the basic model. In the first extension, we consider a scenario where each task can only be performed a given number of times at most. We develop a $\max\{\mu/(\Delta+1), \mu  /2(\sqrt{m}+1)\}$-approximate algorithm for this extension. In the second extension, we relax the assumption that each person can participate in only one task by allowing  individuals to have different load limits. We develop a $\max\{\mu/(4\Delta), \mu/(8\sqrt{f_{\max}m})\}$-approximate algorithm for this extension, where $f_{\max}$ represents the largest number of tasks an individual can participate in.
\item Although we focus on \textsc{TeamGrouping}, our results apply to other applications, such as the lifetime maximization problem in wireless networks \citep{bagaria2013optimally}, resource allocation and scheduling problems \citep{pananjady2014maximizing}, and supply chain management problems \citep{lu2011fundamentals}. In this sense, this research contributes fundamentally  to any problems that fall into the family of generalized cover decomposition problem.
\end{itemize}

The remainder of this paper is organized as follows. In
Section \ref{sec:related}, we review the literature on team formation and disjoint set cover. We introduce the formulation of our problem in Section \ref{sec:system}. In Section \ref{sec:LP}, we present our LP-based approximation algorithms. We conduct extensive experiments in Section \ref{sec:exp}. The two extensions of the basic model are studied in Section \ref{sec:extend}. We summarize this study in Section \ref{sec:conclu}.  Most notations used in this paper are summarized in Table \ref{symbol}.

\begin{table}[ht]\centering
\caption{Symbol table.}
\begin{tabular}{{|c|c|}}
\hline
\textbf{Notation} & \textbf{Meaning}\\
\hline
\hline
$n, m, k$ & Number of skills, individuals, tasks\\
\hline
$\Delta$ & Size of the largest minimal team\\
\hline
$\mathcal{C}$ & Ground set of teams\\
\hline
$\mathcal{C}_t \in \mathcal{C}$ & Set of teams covering task $t$\\
\hline
$C_{ti}\in \mathcal{C}_t$ & $i$th team in $\mathcal{C}_t$\\
\hline
$1/\mu$ & Approximation ratio of the \textsc{MincostTeamSelection} problem\\
\hline
$x^*$ & (Approximate) Solution of primal LP\\
\hline
$\mathcal{N} (C)$ & $C$'s adjacent teams from $\mathcal{C}^I$\\
\hline
$\mathcal{C}(x^*)$ & $\mathcal{C}(x^*)=\{C_{ti}\mid x^*_{ti}>0\}$\\
\hline
$\mathcal{C}(x^*)_t$ & $\mathcal{C}^H_t = \mathcal{C}(x^*)\cap \mathcal{C}_t$\\
\hline
\end{tabular}
\label{symbol}
\end{table}

\section{Related Work}
\label{sec:related}
To the best of our knowledge, we are the first to formulate
and study the team grouping problem and its variants. However, our work is closely
related to other team formation and hiring problems. \cite{lappas2009finding} introduced the minimum cost team formation problem. Given a set of skills to be covered and a social network, the objective is to select a team of experts that can  cover all required skills, while ensuring efficient communication between team members. There is a considerable amount of literature on this topic and its variants \citep{kargar2013finding,dorn2010composing,gajewar2012multi,kargar2011discovering,li2010team,sozio2010community}. \cite{golshan2014profit} studied the cluster hiring problem, where the objective is to hire a profit-maximizing team of experts who can complete multiple projects within a fixed budget. The aforementioned studies aim to  select a single team. By contrast,  our objective is to group individuals into multiple teams. Nevertheless, our problem is closely related to the team formation problem, and we use their solution as a key component of our solution.

Another category of related work is the \emph{maximum disjoint set cover} (DSCP)  problem \citep{bagaria2013optimally}. Given a universe, and a set of subsets, the objective of  this  problem is to find as many set covers as possible such that all set covers are pairwise disjoint. Our problem can be considered a generalization of DSCP because every task in our problem may have different coverage requirements, capacity constraints, and profits. Moreover, every feasible set cover (team) in our problem must satisfy both coverage requirement and social compatibility. In addition, the requirement of ``disjointness'' is relaxed in our problem by allowing individuals to have different load limits. 

\section{Problem Formulation}
\label{sec:system}
\paragraph{\textbf{Individuals. Skills. Tasks.}} Consider
a set of  $n$ skills $\mathcal{S}$, a set of $m$ individuals $\mathcal{V}$, and a set of $k$
tasks $\mathcal{T}$. Each individual $u \in \mathcal{V}$ is represented by a subset
of skills  possessed by this individual; that is, $u \subseteq \mathcal{S}$; these are the skills that the individual
possesses. Every task $t \in \mathcal{T}$ can also be represented
by the set of skills needed to complete the task (i.e., $t \subseteq \mathcal{S}$).
In addition, each task $t\in \mathcal{T}$ has a profit $\lambda_t$. We assume that each task has an unlimited number of copies; that is,
the same task can be performed by an unlimited number of  teams. We relax this assumption in Section \ref{sec:extend} by imposing a capacity constraint on each task.

\paragraph{\textbf{Load.}} Our basic model assumes that each individual can participate in only \emph{one} task. In Section \ref{sec:extend}, we relax this assumption by allowing individuals to have different load limits. 

\paragraph{\textbf{Teams.}} In practice,  social compatibility  between individuals plays an important role in teamwork. For example, low social compatibility  or high coordination costs  might degrade the organizational efficiency \citep{coase1937nature}. We model \emph{social compatibility}  by means of a social network $G=(\mathcal{V}, \mathcal{E})$,  where the nodes in $\mathcal{V}$ represent individuals and an edge in $\mathcal{E}$ connecting two
nodes denotes the social connection between the corresponding individuals. \emph{Connectivity} is a widely known concept that captures the underlying social compatibility
of a team.  This follows the approach of \cite{lappas2009finding} and requires that each team form a connected graph. Another popular indicator of social compatibility is \emph{diameter} \citep{anagnostopoulos2012online}; that is, the longest shortest path between team members in a social network is no longer than a given threshold. Nonetheless, our results are not restricted to any specific notations of social compatibility. 

\paragraph{Problem Formulation.} For a team of individuals $C \subseteq \mathcal{V}$, $C$ is deemed to have
skill $s$ if there exists at least one individual $u \in C$ such that $u$
has skill $s$, that is, $s \in u$. For a task $t \in \mathcal{T}$,
team $C$ is deemed to cover $t$ if $C$ (as a team) has all the skills required
by $t$. A team of individuals may cover more than one
task, but each individual can only participate in one of those tasks \footnote{As mentioned earlier, this assumption will be relaxed in Section \ref{sec:extend}.}.  We define the set of
qualified teams for a task $t\in \mathcal{T}$ to be the set of socially compatible teams covering $t$. That is,
\[\mathcal{C}_t = \{C \subseteq \mathcal{V} \mid C \mbox{ is socially compatible }\wedge C\mbox{ covers } t\}.\]
A minimal team for a task is a qualified team for this task that is not a superset of any other qualified team. In the rest of this paper, we only consider minimal teams. Let $\mathcal{C}=\cup_{t\in \mathcal{T}} \mathcal{C}_t$.
The objective of the \textsc{TeamGrouping} problem is to select a group of teams from $\mathcal{C}$ such that each individual participates in only one team. We formally define the \textsc{TeamGrouping} problem in \textbf{P.1}.
For each $t\in \mathcal{T}$ and $i\in \{0, 1, \cdots, |\mathcal{C}_t|\}$, let $C_{ti}$ denote the $i$th team in $\mathcal{C}_t $.  Let $ x_{ti}$ be an indicator of whether team $C_{ti}$ is selected ($ x_{ti}=1$) or not ($ x_{ti}=0$).

\begin{center}
\framebox[0.78\textwidth][c]{
\enspace
\begin{minipage}[t]{0.45\textwidth}
\small
\textbf{P.1:}
\emph{Maximize $ \sum_{C_{ti}\in \mathcal{C}}(x_{ti}\cdot \lambda_t)$}\\
\textbf{subject to:}
\begin{equation*}
\begin{cases}
 \sum_{C_{ti}\in \mathcal{C}:  C_{ti} \ni u} x_{ti} \leq 1, \forall u\in \mathcal{V} \\
 x_{ti} \in \{0,1\}, \forall C_{ti}\in \mathcal{C}.

\end{cases}
\end{equation*}
\end{minipage}
}
\end{center}
\vspace{0.1in}

The first constraint specifies that each individual participates in at most one team.
Recall that $|\mathcal{V}|=m$, the following results show that we cannot hope to achieve an  approximation ratio of $\omega(1/\ln m)$ for this problem.

\begin{theorem}
Let $m=|\mathcal{V}|$. \textbf{P.1} is $1/\ln m$-hard to approximate.
\end{theorem}
\emph{Proof:} For this proof, we consider a simplified version of  \textbf{P.1}. There is only one task, that is, $k=1$, and there is no constraint on social compatibility. We call this problem \textsc{s-TeamGrouping}. We next prove that the DSCP can be reduced to \textsc{s-TeamGrouping}.
The formal definition of DSCP is as follows: Given a universe $\mathcal{U}$ and a set of subsets $\mathcal{X}$, the goal is to find as many set covers as possible
such that all set covers are pairwise disjoint. We wish to formulate an equivalent \textsc{s-TeamGrouping} with a set of skills $\mathcal{S}$ required to do the task, and a set of individuals $\mathcal{V}$. Let $\mathcal{S}=\mathcal{U}$ and $\mathcal{V}=\mathcal{X}$.  Because there is only one task and no constraint on social compatibility, \textsc{s-TeamGrouping} is equivalent to grouping $\mathcal{V}$ into the maximum number of disjoint teams such that each team can cover all skills in $\mathcal{S}$. According to \cite{bagaria2013optimally}, it is hard to achieve an  approximation ratio of $\omega(1/\ln m)$ unless $NP \subseteq DTIME (n^{O(\ln\ln m)})$. Thus,  \textbf{P.1}, which is a generalization of \textsc{s-TeamGrouping}, is also $1/\ln m$-hard to approximate. $\Box$

\cite{bagaria2013optimally} developed an $1/\ln m$-approximate algorithm for DSCP. For the special case of our problem where there is only one task and no constraint on social compatibility, we can simply adopt their method to achieve an  approximation ratio of $1/\ln m$. In the following, we propose an LP-based approximation algorithm to address the general case.
\section{LP-Based Approximation Algorithms}
\label{sec:LP}
In this section, we give a $\max\{\mu/\Delta, \mu  /2\sqrt{m}\}$-approximation algorithm for  \textbf{P.1}, where $1/\mu$ is the approximation factor of the algorithm for the \textsc{MincostTeamSelection} problem, which is formally defined in Definition \ref{def:min}, and $\Delta:=\max_{C\in \mathcal{C}} |C|$ is the size of the largest minimal team. Our algorithm consists of two phases: we first solve the LP relaxation of the original problem  to obtain a fractional solution (Section \ref{sec:LP Relaxation}) and then use this fractional solution to compute a group of teams (Section \ref{sec:approxi}).

\subsection{LP Relaxation}
\label{sec:LP Relaxation}
We first present the LP relaxation of \textbf{P.1}. We refer to this relaxation as
the primal LP.
\begin{center}
\framebox[0.78\textwidth][c]{
\enspace
\begin{minipage}[t]{0.55\textwidth}
\small
\textbf{Primal LP of P.1:}
\emph{Maximize $ \sum_{C_{ti}\in \mathcal{C}}(x_{ti}\cdot \lambda_t)$}\\
\textbf{subject to:}
\begin{equation*}
\begin{cases}
\sum_{C_{ti}\in \mathcal{C}:  C_{ti} \ni u} x_{ti} \leq 1, \forall u\in \mathcal{V} \\
 x_{ti}\geq 0, \forall C_{ti}\in \mathcal{C}.

\end{cases}
\end{equation*}
\end{minipage}
}
\end{center}
\vspace{0.1in}

This LP has $m$ constraints (excluding the trivial
constraints $x_{ti} \geq 0$). However, its number of variables is
$\sum_{t\in \mathcal{T}} |\mathcal{C}_t|$, which can easily be exponential in the number of
individuals. Hence, standard LP solvers cannot solve this packing LP effectively.

To address this challenge, we rely on the ellipsoid algorithm \citep{grotschel1981ellipsoid} and the dual problem (\textbf{Dual LP of P.1}). 
On a high level, we use the ellipsoid method to test whether a given non-degenerate convex set $S$ is empty or not. Here, $S$ represents the feasibility region of the dual problem. This method starts with an ellipsoid that is guaranteed to contain $S$. In each iteration, it determines whether the center of the current ellipsoid is in $S$. If the answer is ``yes,'', then $S$ is nonempty, which indicates that the current solution is feasible. In this case, the method
tries a smaller ellipsoid that decreases the objective function. Otherwise, the method finds a violated constraint through  an (approximate) separation oracle and tries a smaller ellipsoid whose center satisfies that constraint. Geometrically,  we take a hyperplane through the center of the original ellipsoid such that $S$ is contained in one  of the two half-ellipsoids. We take the smallest ellipsoid completely containing this half-ellipsoid, whose volume is
substantially smaller than the volume of the previous ellipsoid. This process iterates until the volume of the bounding ellipsoid is sufficiently small, in which case $S$ is considered empty; that is, we cannot find a feasible solution with a smaller objective. This process takes  a polynomial number of iterations for solving linear problems. We do not require an explicit description of LP to make this method work; we only need a polynomial-time (approximate) separation oracle to examine whether a point lies
in $S$ or not and,  in the latter case, return a separating hyperplane.

Here, we formally introduce our algorithm.  We next present \textbf{Dual LP of P.1}, the dual to the primal LP.  In the dual problem, we assign a
price $y(u)$ to each node $u \in \mathcal{V}$.
\begin{center}
\framebox[0.78\textwidth][c]{
\enspace
\begin{minipage}[t]{0.45\textwidth}
\small
\textbf{Dual LP of P.1:} \emph{Minimize $\sum_{u\in \mathcal{V}} y(u)$}\\
\textbf{subject to:}
\begin{equation*}
\begin{cases}
 \sum_{u\in C_{ti}} y(u) \geq \lambda_t, \forall C_{ti}\in \mathcal{C}\\
y(u) \geq 0, \forall u \in \mathcal{V}.

\end{cases}
\end{equation*}
\end{minipage}
}
\end{center}
\vspace{0.1in}

We leverage the ellipsoid method for exponential-sized
LP with an (approximate) separation
oracle to solve this problem. In particular, in each iteration of the ellipsoid method, we solve the
 \textsc{MincostTeamSelection} problem approximately to obtain a polynomial-time approximate separation oracle to check the feasibility of the current solution.

\begin{definition}[MincostTeamSelection]
\label{def:min}
Assume that there is
a set of  skills $\mathcal{S}$ and individuals $\mathcal{V}$; each individual $u\in \mathcal{V}$ has a cost and possesses a subset of skills. We indentify a team of individuals with the minimum cost such that (1) all team members are socially compatible, and (2) all skills in $\mathcal{S}$ can be covered.
\end{definition}

   \textsc{MincostTeamSelection} has been intensively studied in the literature, using various indicators of social compatibility. For example, if there is no requirement of social compatibility, then  \textsc{MincostTeamSelection} reduces to the classical \emph{weighted set cover problem}  \citep{chvatal1979greedy}, which admits  an $O(\log n)$-factor approximation. \citep{lappas2009finding} proposed the use of  connectivity as a measure of social compatibility; that is, all team members must be connected in a social network. In this context, the \textsc{MincostTeamSelection} problem  can be reduced from \emph{node weight group steiner tree} problem \citep{khandekar2012approximating}, which admits a performance ratio of $O(|\mathcal{E}|^{1/2} \ln |\mathcal{E}|)$, where $|\mathcal{E}|$ is the number of edges in the social network. As stated by \cite{anagnostopoulos2012online}, a team must have a bounded diameter.  
  We next present the main theorem of this section. This theorem is not restricted to any specific indicator of social compatibility.
\begin{theorem}
\label{thm:2}
If there is a polynomial $1/\mu$-approximation algorithm for \textsc{MincostTeamSelection}, then there exists a polynomial $\mu$-approximation algorithm for \textbf{Primal LP of P.1}.
\end{theorem}
\emph{Proof:} Let $\mathcal{A}$ be a $1/\mu$-approximation algorithm for
\textsc{MincostTeamSelection}. We use $\mathcal{A}$ as an approximate separation
oracle to examine whether the current solution to the dual problem  is feasible or not. Let $S (L)$ denote the set of $y\in \mathbb{R}_+^\mathcal{V}$ satisfying that
\[\sum_{u\in \mathcal{V}} y(u) \leq L,\]
\[\sum_{u\in C_{ti}} y(u) \geq \lambda_t, \forall C_{ti}\in \mathcal{C}.\]

We implement binary search to find the smallest value of $L$ for
which $S (L)$ is nonempty. For a given $L$, the method first checks the inequality $\sum_{u\in \mathcal{V}} y(u) \leq L$. Then, it runs algorithm $\mathcal{A}$, using $y(u)$ as the price function to select the cheapest group $C_{t} \in \mathcal{C}_t$ for each task $ t\in \mathcal{T}$. Suppose $\mathcal{A}$ is an exact algorithm, that is, $\mu=1$. If for all $t$, $\sum_{u\in C_{t}} y(u) \geq \lambda_t$, then $y\in S(L)$. If there exists some $t$ such that $\sum_{u\in C_{t}} y(u) < \lambda_t$, then $y\notin S(L)$ and $C_{t}$ is a separating hyperplane. However, for general $\mu\leq 1$, $C_{t} \in \mathcal{C}_t$ might not be the cheapest team for task $ t\in \mathcal{T}$. Hence, $S (L)$ might actually be empty even if  $\forall t, \sum_{u\in C_{t}} y(u) \geq \lambda_t$. Nonetheless, even for this general case,
$\frac{1}{\mu} \cdot y \in S (\frac{1}{\mu} \cdot  L)$. Let $L^*$ be the minimum value
of $L$ for which the algorithm decides $S (L)$ is nonempty. We can conclude that $S (\frac{1}{\mu} \cdot L^*)$ is nonempty and  $S (L^* - \epsilon)$ is empty, where $\epsilon$ is the precision of the algorithm. That is, the value of the dual LP and thus
the value of the primal LP belong to $[L^*-\epsilon,\frac{1}{\mu} \cdot L^*]$. Therefore, by finding a solution of value $L^*-\epsilon$  for the primal LP, we achieve an approximation ratio of $\mu$ against the optimal solution.

Here, we explain how to  compute such a  solution  using only teams corresponding to the separating hyperplanes found by the separation oracle. Let $\mathcal{C}^H_t$ denote the subset of teams in $\mathcal{C}_t$  for which the dual constraint is violated
in the implementation of the ellipsoid algorithm on $S (L^* - \epsilon)$.
Then, $\sum_{t=1}^{k} |\mathcal{C}^H_t|$ is polynomial. Let $\mathcal{C}^H = \cup_{t\in \mathcal{T}} \mathcal{C}^H_t$,  and consider the restricted dual LP.

\begin{center}
\enspace
\begin{minipage}[t]{0.55\textwidth}
\small
 \emph{Minimize $\sum_{u\in \mathcal{V}} y(u)$}\\
\textbf{subject to:}
\begin{equation*}
\begin{cases}
 \sum_{u\in C_{ti}} y(u) \geq \lambda_t, \forall C_{ti} \in \mathcal{C}^H\\
y(u) \geq 0, \forall u \in \mathcal{V}.
\end{cases}
\end{equation*}
\end{minipage}
\end{center}
\vspace{0.1in}

The value of the optimal solution to the above restricted dual LP is also at least $L^*$. Thus, we solve the following
restricted primal LP of polynomial size, which is the
dual of the restricted dual LP:
\begin{center}
\enspace
\begin{minipage}[t]{0.45\textwidth}
\small
\emph{Maximize $\sum_{C_{ti}\in \mathcal{C}^H }(x_{ti}\cdot \lambda_t)$}\\
\textbf{subject to:}
\begin{equation*}
\begin{cases}
 \sum_{C_{ti}\in \mathcal{C}^H: u\ni  C_{ti}} x_{ti} \leq 1, \forall u\in \mathcal{V}\\
 x_{ti}\geq 0,  \forall C_{ti} \in \mathcal{C}^H.

\end{cases}
\end{equation*}
\end{minipage}
\end{center}
\vspace{0.1in}
The value of the optimal solution of this restricted LP is
at least $L^*$, which is a $\mu$-approximation to the original
primal LP. $\Box$

\subsection{Approximation Algorithm}
\label{sec:approxi}
Before presenting our algorithm, we present a deterministic rounding method that converts any feasible solution of \textbf{Primal LP of P.1} to a feasible solution of \textbf{P.1}. Later, we use this rounding method as an essential subroutine to build our final algorithm.

\subsubsection{LP Rounding}
\label{sec:round}
 Given any feasible solution $x^*=\{x^*_{ij}\mid C_{ti}\in \mathcal{C}\}$ of \textbf{Primal LP of P.1}, let $\mathcal{C}(x^*)=\{C_{ti}\mid x^*_{ti}>0\}$ denote the set of all teams whose fractional value in $x^*$ is positive. Two teams are considered \emph{adjacent} if they contain at least one common individual. $\mathcal{N}(C, \mathcal{C}^{I})$ denotes the set of all adjacent teams of $C$ from a set of input teams $\mathcal{C}^{I}\subseteq \mathcal{C}(x^*)$, that is, $\mathcal{N}(C, \mathcal{C}^{I})=\{C'\in \mathcal{C}^{I}\mid C'\neq C\wedge C\cap C' \neq \emptyset\}$. For simplicity, we use  $\mathcal{N}(C)$ to denote  $\mathcal{N}(C, \mathcal{C}^{I})$ when it is clear from the context.

 Our deterministic rounding method (Algorithm \ref{alg:LPP0}) takes a set of teams $\mathcal{C}^{I}\subseteq \mathcal{C}(x^*)$ as input.

\emph{Step 1:} Select the team that has the highest profit from $\mathcal{C}^{I}$ (e.g., $C_{ti}$,).

\emph{Step 2:} Add $C_{ti}$ to $\mathcal{C}^{DR}$ and remove $C_{ti}\cup \mathcal{N}(C_{ti})$ from $\mathcal{C}^{I}$.  This step ensures that no individual participates in multiple tasks. Go to Step 1 unless there are no teams left. Output $\mathcal{C}^{DR}$.

\begin{algorithm}[hptb]
\caption{Deterministic Rounding}
\label{alg:LPP0}
\textbf{Input}: $\mathcal{C}^{I}\subseteq \mathcal{C}(x^*)$.
\begin{algorithmic}[1]
\STATE $\mathcal{C}^{DR}=\emptyset$
\WHILE {$\mathcal{C}^{I}  \neq \emptyset$}
\STATE Select the team, say $C_{ti}$, that has the highest profit from $\mathcal{C}^{I}$.
\STATE $\mathcal{C}^{DR} = \mathcal{C}^{DR} \cup \{C_{ti}\}$.
\STATE $\mathcal{C}^{I}=\mathcal{C}^{I} \setminus \{C_{ti}\cup \mathcal{N}(C_{ti})\}$.
\ENDWHILE
\STATE Return $\mathcal{C}^{DR}$.
\end{algorithmic}
\end{algorithm}

Let $\rho(\mathcal{C}^{I})=\max_{C_{ti}\in \mathcal{C}^{I}}|C_{ti}|$ denote the size of the largest team in $\mathcal{C}^{I}$. We next show that the profit of $\mathcal{C}^{DR}$ is at least $1/\rho(\mathcal{C}^{I})$ faction of the one obtained from the fractional solution $x^*$.

\begin{lemma}
\label{the:constant}
Given a feasible solution $x^*$ of \textbf{Primal LP of P.1}, a set of input teams $\mathcal{C}^{I}\subseteq \mathcal{C}(x^*)$, $\sum_{C_{ti} \in \mathcal{C}^{DR}} \lambda_t \geq \sum_{C_{ti} \in \mathcal{C}^{I}}(x^*_{ti}\cdot \lambda_t)/\rho(\mathcal{C}^{I})$, where $\rho(\mathcal{C}^{I})=\max_{C_{ti}\in \mathcal{C}^{I}}|C_{ti}|$.
\end{lemma}
\emph{Proof:} Consider any team $C_{ti} \in \mathcal{C}^{DR}$. We have \begin{eqnarray}
&&x^*_{ti}\cdot \lambda_t + \sum_{C_{lj}\in \mathcal{N}(C_{ti})} (x^*_{lj}\cdot \lambda_l) \leq x^*_{ti}\cdot \lambda_t + \sum_{C_{lj}\in \mathcal{N}(C_{ti})} (x^*_{lj}\cdot \lambda_t) \\
&&=  \lambda_t\times ( x^*_{ti} + \sum_{C_{lj}\in \mathcal{N}(C_{ti})} x^*_{lj}) \\
&&\leq  \lambda_t\times \sum_{u\in C_{ti}}\sum_{C_{lj}\in C_{ti}\cup\mathcal{N}(C_{ti}): C_{lj}\ni u} x^*_{lj}\\
&&\leq  \lambda_t\times \sum_{u\in C_{ti}}\sum_{C_{lj}\in \mathcal{C}(x^*): C_{lj}\ni u} x^*_{lj}\\
&&\leq  \lambda_t\times \sum_{u\in C_{ti}}1\\
&&\leq \rho(\mathcal{C}^{I}) \cdot \lambda_t. \label{eq:aaai}
 \end{eqnarray}
 The first inequality is due to $C_{ti}$ having the highest profit among all its adjacent teams; the second inequality is due to the definition of  $\mathcal{N}(C_{ti})$; the fourth inequality is due to $x^*$ being a feasible solution of \textbf{Primal LP of P.1}, indicating that $\sum_{C_{lj}\in \mathcal{C}(x^*): C_{lj}\ni u} x^*_{lj} \leq 1, \forall u\in C_{ti}$; the last inequality is due to $\rho(\mathcal{C}^{I})=\max_{C_{ti}\in \mathcal{C}^{I}}|C_{ti}|$.

Therefore, for any $C_{ti} \in \mathcal{C}^{DR}$, $\lambda_t \geq (x^*_{ti}\cdot \lambda_t + \sum_{C_{lj}\in \mathcal{N}(C_{ti})} (x^*_{lj}\cdot \lambda_l))/\rho(\mathcal{C}^{I})$. Summation of  this inequality over all teams from $\mathcal{C}^{DR}$ gives  
\begin{eqnarray}
\sum_{C_{ti} \in \mathcal{C}^{DR}} \lambda_t \geq \sum_{C_{ti} \in \mathcal{C}^{DR}}(x^*_{ti}\cdot \lambda_t + \sum_{C_{lj}\in \mathcal{N}(C_{ti})} (x^*_{lj}\cdot \lambda_l))/\rho(\mathcal{C}^{I}) = \sum_{C_{ti} \in \mathcal{C}^{I}}(x^*_{ti}\cdot \lambda_t)/\rho(\mathcal{C}^{I}).\end{eqnarray} That is, the profit of $\mathcal{C}^{DR}$ is at least $1/\rho(\mathcal{C}^{I})$ of the one obtained from the fractional solution $x^*$. $\Box$

\subsection{Algorithm Design and Performance Analysis}
\label{sec:444}
 We present our final algorithm, \textsf{Approx-TG}.
 In the rest of this paper, $OPT$ denotes the profit gained from the optimal grouping. \textsf{Approx-TG} selects the better solution from two candidates as the final output. We present these two candidate solutions in detail.

\textbf{Candidate Solution I:} In Algorithm \ref{alg:LPP1}, we directly apply the deterministic rounding (Algorithm \ref{alg:LPP0}) to $\mathcal{C}(x^*)$, that is, we feed $\mathcal{C}^{I}=\mathcal{C}(x^*)$ as input teams to Algorithm \ref{alg:LPP0}. We prove that if $x^*$ is a $\mu$-approximate solution of \textbf{Primal LP of P.1}  that is found by the ellipsoid method, then Algorithm \ref{alg:LPP1} achieves an approximation ratio  of $\mu/\Delta$,  where $\Delta$ denotes the size of the largest minimal team.

\begin{algorithm}[hptb]
\caption{Candidate Grouping -  I}
\label{alg:LPP1}
\textbf{Input}: $x^*$.
\begin{algorithmic}[1]
\STATE Apply deterministic rounding (Algorithm \ref{alg:LPP0}) to $x^*$ and output a group of teams.
\end{algorithmic}
\end{algorithm}

\begin{lemma}
\label{lem:1}
Assume that $x^*$ is a $\mu$-approximate solution of \textbf{Primal LP of P.1} that is found by the ellipsoid method. Then Algorithm \ref{alg:LPP1}  achieves an approximation ratio of $\mu/\Delta$  for \textbf{P.1}.
\end{lemma}
\emph{Proof:} By Lemma \ref{the:constant}, Algorithm \ref{alg:LPP0} takes $\mathcal{C}(x^*)$ as input and returns a grouping that achieves a profit of at least $\frac{1}{\Delta}  \sum_{C_{ti}\in \mathcal{C}(x^*)} (x^*_{ti}\cdot \lambda_t)$, where  $\Delta$ is the size of the largest possible team in $\mathcal{C}(x^*)$. By the assumption that  $x^*$ is a $\mu$-approximate solution of \textbf{Primal LP of P.1}, we have $\sum_{C_{ti}\in \mathcal{C}(x^*)} (x^*_{ti}\cdot \lambda_t)\geq \mu\cdot OPT$. Thus, Algorithm \ref{alg:LPP1} achieves a profit of at least
\[\frac{1}{\Delta}  \sum_{C_{ti}\in \mathcal{C}(x^*)} (x^*_{ti}\cdot \lambda_t)\geq \frac{\mu}{\Delta}\cdot OPT.\]
 $\Box$

\textbf{Candidate Solution II:} Let $\mathcal{C}(x^*)_t=\mathcal{C}(x^*)\cap \mathcal{C}_t$ denote the subset of $\mathcal{C}(x^*)$ that is assigned to   task $t\in \mathcal{T}$. Hence, $\mathcal{C}(x^*)=\cup_{t\in \mathcal{T}}\mathcal{C}(x^*)_t$. The framework of the second candidate solution (Algorithm \ref{alg:LPP2}) is summarized as follows:

\emph{Step 1:}  For every task $t\in \mathcal{T}$, we first partition $\mathcal{C}(x^*)_t$ into the disjoint subsets $\mathcal{C}(x^*)^1_t$ and $\mathcal{C}(x^*)^2_t$ such that $\forall C \in \mathcal{C}(x^*)^1_t: |C|\leq \sqrt{m}$ and $\forall C \in \mathcal{C}(x^*)^2_t: |C| > \sqrt{m}$. That is, $\mathcal{C}(x^*)^1_t$ (resp. $\mathcal{C}(x^*)^2_t$) contains all teams with no more (resp. less) than $\sqrt{m}$ individuals. Let $\mathcal{C}(x^*)^1 = \cup_{t\in \mathcal{T}}  \mathcal{C}(x^*)^1_t$ and $\mathcal{C}(x^*)^2 = \cup_{t\in \mathcal{T}} \mathcal{C}(x^*)^2_t$.

\emph{Step 2:} Apply deterministic rounding (Algorithm \ref{alg:LPP0}) to $\mathcal{C}(x^*)^1$ to obtain a group of teams $\widetilde{\mathcal{C}}$.

\emph{Step 3:} Select a team from $\mathcal{C}(x^*)^2$ whose task $t_{\max}$ has the highest profit $\lambda_{t_{\max}}$ (e.g., $C_{t_{\max}}$).

\emph{Step 4:} Output the better solution between $\widetilde{\mathcal{C}}$ and $\{C_{t_{\max}}\}$ as the final output; that is, the profit of the returned solution is $\max\{\sum_{C_{ti}\in \widetilde{\mathcal{C}}} \lambda_t, \lambda_{t_{\max}}\}$.

\begin{algorithm}[hptb]
\caption{Candidate Grouping -  II}
\label{alg:LPP2}
\textbf{Input}: $x^*$.
\begin{algorithmic}[1]
\STATE Partition $\mathcal{C}(x^*)$ into two subsets $\mathcal{C}(x^*)^1$ and $\mathcal{C}(x^*)^2$.
\STATE Apply the deterministic rounding (Algorithm \ref{alg:LPP0}) to $\mathcal{C}(x^*)^1$ to obtain $\widetilde{\mathcal{C}}$.
\STATE Select a team with the highest profit, say $C_{t_{\max}}$, from $\mathcal{C}(x^*)^2$.
\STATE Compare $\widetilde{\mathcal{C}}$ and $\{C_{t_{\max}}\}$, return the one with larger profit.
\end{algorithmic}
\end{algorithm}

We next prove that if $x^*$ is a $\mu$-approximate solution of \textbf{Primal LP of P.1} that is found by the ellipsoid method, then the approximation ratio of Algorithm \ref{alg:LPP2} can be bounded by $\mu  /(2\sqrt{m})$.
\begin{lemma}
\label{lem:2}
Assume that $x^*$ is a $\mu$-approximate solution of \textbf{Primal LP of P.1} that is found by the ellipsoid method. Algorithm \ref{alg:LPP2} achieves an approximation ratio  of $\mu  /(2\sqrt{m})$  for \textbf{P.1}.
\end{lemma}
\emph{Proof:}
To prove this lemma, we show that $\max\{\sum_{C_{ti}\in \widetilde{\mathcal{C}}} \lambda_t, \lambda_{t_{\max}}\}\geq \frac{1}{\sqrt{m}}\cdot \frac{\mu }{2}\cdot OPT$.

We first bound the gap between the profit gained from $\widetilde{\mathcal{C}}$  and $\sum_{C_{ti}\in  \mathcal{C}(x^*)^1}(x^*_{ti}\cdot \lambda_t)$. By Lemma \ref{the:constant}, we have
\begin{equation}
\label{eq:1}\sum_{C_{ti}\in \widetilde{\mathcal{C}}} \lambda_t \geq \frac{1}{\rho}\cdot \sum_{C_{ti} \in \mathcal{C}(x^*)^1}(x^*_{ti}\cdot \lambda_t ) \geq \frac{1}{\sqrt{m}}\cdot \sum_{C_{ti}\in \mathcal{C}(x^*)^1}(x^*_{ti}\cdot \lambda_t ), \end{equation}
 where the second inequality is due to the assumption that $\rho \leq \sqrt{m}$ holds for all teams from $\mathcal{C}(x^*)^1$.

We next bound the gap between the profit gained from $C_{t_{\max}}$ and $\sum_{C_{ti}\in  \mathcal{C}(x^*)^2}(x^*_{ti}\cdot \lambda_t)$. In particular, we show that
\begin{eqnarray}
\label{eq:211}
\lambda_{t_{\max}}\geq \frac{1}{\sqrt{m}} \cdot \sum_{C_{ti} \in \mathcal{C}(x^*)^2}(x^*_{ti}\cdot \lambda_t).
\end{eqnarray}

The following chain proves this inequality:
\begin{equation}
\label{eq:2}\sum_{C_{ti} \in \mathcal{C}(x^*)^2}(x^*_{ti}\cdot \lambda_t) \leq \sum_{C_{ti} \in \mathcal{C}(x^*)^2}(x^*_{ti}\cdot \lambda_{t_{\max}})=\lambda_{t_{\max}}\cdot \sum_{C_{ti} \in \mathcal{C}(x^*)^2}x^*_{ti}\leq \lambda_{t_{\max}}\cdot \frac{m}{\sqrt{m}}.
\end{equation}
The first inequality is due to the assumption that $C_{t_{\max}}$ delivers the highest profit among $\mathcal{C}(x^*)^2$. We then prove the second inequality. Because  $x^*$ is a feasible solution of \textbf{Primal LP of P.1} and $\mathcal{C}(x^*)^2\subseteq \mathcal{C}$, we have $\sum_{C_{ti}\in \mathcal{C}(x^*)^2:  C_{ti} \ni u} x^*_{ti} \leq 1, \forall u\in \mathcal{V}$. Therefore,
\begin{eqnarray}
\label{eq:rain}
\sum_{C_{ti} \in \mathcal{C}(x^*)^2}(x^*_{ti}\cdot |C_{ti}|) \leq m.
 \end{eqnarray}

 All teams in $\mathcal{C}(x^*)^2$ contain at least $\sqrt{m}$ individuals, so $\sum_{C_{ti} \in \mathcal{C}(x^*)^2}(x^*_{ti}\cdot |C_{ti}|) \geq \sum_{C_{ti} \in \mathcal{C}(x^*)^2}(x^*_{ti}\cdot \sqrt{m})$. This, together with (\ref{eq:rain}), implies that
$\sum_{C_{ti} \in \mathcal{C}(x^*)^2}(x^*_{ti}\cdot \sqrt{m}) \leq m$; thus, $\sum_{C_{ti}\in \mathcal{C}(x^*)^2}(x^*_{ti}) \leq \sqrt{m}$. This finishes the proof of the second inequality.

By the assumption that $x^*$ is a $\mu$-approximate solution of \textbf{Primal LP of P.1}, we have
\begin{equation}
\label{eq:3}\sum_{C_{ti} \in \mathcal{C}(x^*)}(x^*_{ti}\cdot \lambda_t) \geq \mu \cdot OPT \Rightarrow \sum_{C_{ti} \in \mathcal{C}(x^*)^1}(x^*_{ti} \cdot \lambda_t) + \sum_{C_{ti} \in \mathcal{C}(x^*)^2}(x^*_{ti} \cdot \lambda_t) \geq \mu \cdot OPT.\end{equation}

We now prove this theorem.
\begin{eqnarray}
&&\max\{\sum_{C_{ti}\in \widetilde{\mathcal{C}}} \lambda_t, \lambda_{t_{\max}}\}\geq \frac{\sum_{C_{ti}\in \widetilde{\mathcal{C}}} \lambda_t+ \lambda_{t_{\max}}}{2}\\
&&\geq \frac{1}{\sqrt{m}}\cdot \frac{ \sum_{C_{ti} \in \mathcal{C}(x^*)^1}(x^*_{ti} \cdot \lambda_t) + \sum_{C_{ti} \in \mathcal{C}(x^*)^2}(x^*_{ti}\cdot \lambda_t) }{2}\\
&&\geq \frac{1}{\sqrt{m}}\cdot \frac{\mu }{2}\cdot OPT
\end{eqnarray}
The second inequality is due to (\ref{eq:1}) and (\ref{eq:211}), and the third inequality is due to  (\ref{eq:3}).
$\Box$


\textbf{Putting It All Together.}
Given solutions returned from Algorithms \ref{alg:LPP1} and \ref{alg:LPP2}, \textsf{Approx-TG} returns the one with the higher profit as the final output. 
Lemmas \ref{lem:1} and  \ref{lem:2} jointly imply our main theorem.
\begin{theorem}
\label{thm:main}
\textsf{Approx-TG} achieves an  approximation ratio of $\max\{\mu/\Delta, \mu  /2\sqrt{m}\}$ for \textbf{P.1}.
\end{theorem}

Consider a special case of \textsc{TeamGrouping} where there is no requirement of social compatibility. In this case, the \textsc{MincostTeamSelection} problem reduces to the classical \emph{weighted set cover problem}, which admits an $\ln n$ approximation. In addition, $\Delta \leq n$, because the number of possible skills is at most $n$, and if there is no constraint on social compatibility, then any  minimal team contains at most $n$ individuals. Corollary \ref{cor:6} holds by replacing $\mu$ with $\ln n$, and $\Delta$ with $n$ in Theorem \ref{thm:main}.
\begin{corollary}
\label{cor:6}
If there is no constraint on social compatibility, Approx-TG achieves an approximation ratio of $\max\{\ln n/n, \ln n /2\sqrt{m}\}$ for \textbf{P.1}.
 \end{corollary}

  In practice, $n \ll m$; that is, the number of skills is much smaller than the number of individuals, so the above approximation ratio can be further rewritten as $\ln n/n$.

Consider a special case that uses connectivity as an indicator of social compatibility. As discussed in Section \ref{sec:LP Relaxation}, in this setting, the \textsc{MincostTeamSelection} problem  reduces to a \emph{node weight group Steiner tree} problem \citep{khandekar2012approximating}, which admits a performance ratio of $O(|\mathcal{E}|^{1/2} \ln |\mathcal{E}|)$. Therefore, we have Corollary \ref{cor:2}.

\begin{corollary}
\label{cor:2}
If all teams are required to be connected, Approx-TG achieves an approximation ratio of  $\max\{O(|\mathcal{E}|^{1/2} \ln |\mathcal{E}|)/\Delta), O(|\mathcal{E}|^{1/2} \ln |\mathcal{E}|)/2\sqrt{m})\}$ for \textbf{P.1}.
 \end{corollary}
	\section{Performance Evaluation}
\label{sec:exp}
	In this section, we conduct simulations to evaluate the performance of our algorithm. All experiments were run $10$ times on a desktop with Intel(R) Xeon(R) Gold 5218R CPU @ 2.1GHz and 94GB memory, running 64-bit Linux server. We show that our algorithm outperforms three benchmarks, and we also validate its robustness under various settings.
	
\subsection{Setting}
The input of our basic setting is composed of a set of 10 skills, a set of 20 tasks and a set of 100 individuals. We set the profit of each task to
\[\# \mbox{ skills required by a task }\times r,\]
where $r$ is a random number chosen from $\{1,2,3\}$. We consider two scenarios as follows:

Scenario 1: In the first scenario, we assume there is no constraint on social compatibility. Hence, \textsc{MincostTeamSelection} reduces to the weighted set cover problem. We use greedy algorithm  \citep{slavik1997tight} to solve this problem to obtain a  $\ln n$-approximation solution, where $n$ is the number of elements to be covered.
	
Scenario 2: In the second scenario, we incorporate the constraint of social compatibility. In particular, we use \emph{connectivity} \citep{lappas2009finding} as an indicator of social compatibility; therefore,  each team must form a connected graph.  We generate a  random network consisting of $1000$ edges in the basic setting such that the connecting density of this network is $0.202$. We add more edges to the network in the robustness section. In this scenario,  our \textsc{MincostTeamSelection} problem reduces to the group steiner tree problem, and we use the ImprovAPP algorithm from \citep{sun2021finding} to solve this problem. It has been shown that this algorithm achieves a $(|\Gamma|-1)$-approximation ratio, where $|\Gamma|$ denotes the number of vertex groups.
	\subsection{Benchmark}
We compare our algorithm with three heuristics.
	
\begin{itemize}	
\item \textsf{Random}: 
In each iteration, \textsf{Random} selects a random task, and then builds up a team randomly to cover this task. We remove all selected individuals from consideration in the subsequent iterations. 
This process continues until the individual pool is exhausted or the remaining individuals can not cover any of the tasks.
	
\item \textsf{Greedy}:  \textsf{Greedy} first sorts all tasks in non-increasing order of their profits. Then starting with the first task (e.g., $t$), \textsf{Greedy}  selects a group of teams that covers $t$ sequentially, where each team is selected by solving a weighted set cover problem (resp. the group steiner tree problem) using the greedy algorithm (resp. the ImprovAPP algorithm) in the basic setting (resp. the general setting). If we can not find more teams to cover $t$, then we move to the next task in the list. This process iterates until the individual pool is exhausted or we can not find more teams to cover the last task  in the list.
	
\item \textsf{Greedy+}: Unlike \textsf{Greedy} which ranks tasks according to their profits, \textsf{Greedy+}  ranks tasks according to  $\frac{\lambda_{t}}{|t|}$, the ratio of profit and the number of skills required by a task $t$. The rest of the procedure is identical to \textsf{Greedy}.
\end{itemize}
\subsection{Results}
\begin{figure}[h!]
		\centering
		\includegraphics[scale=0.6]{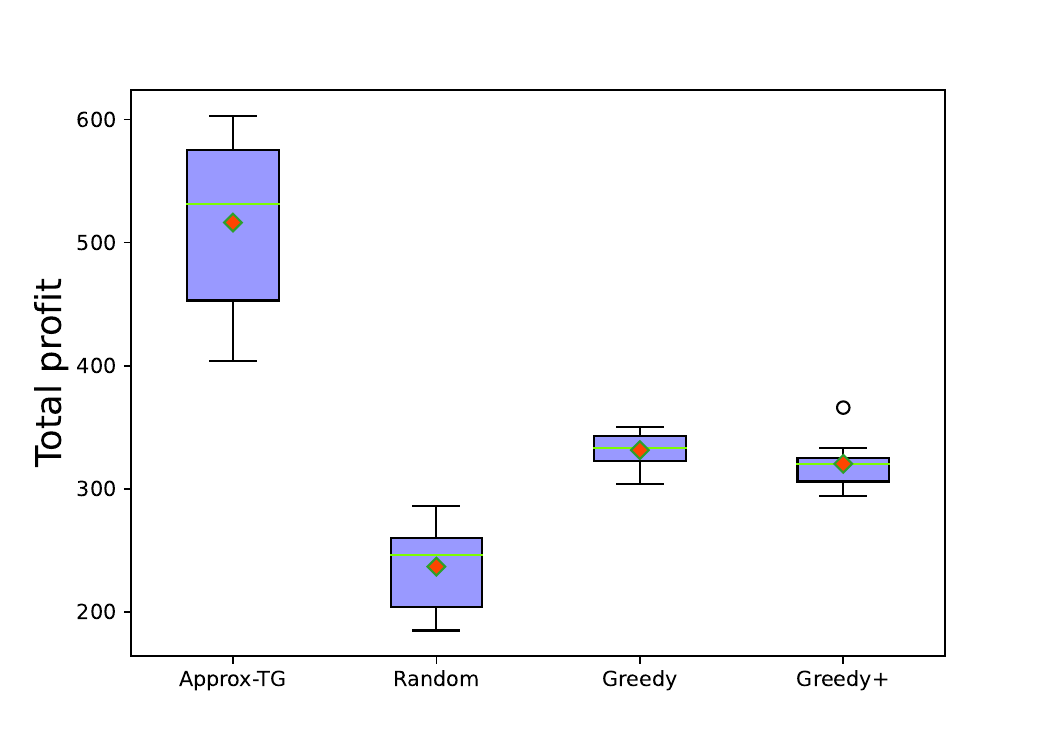}
		\caption{without social compatibility}
		\label{F1}
		\includegraphics[scale=0.6]{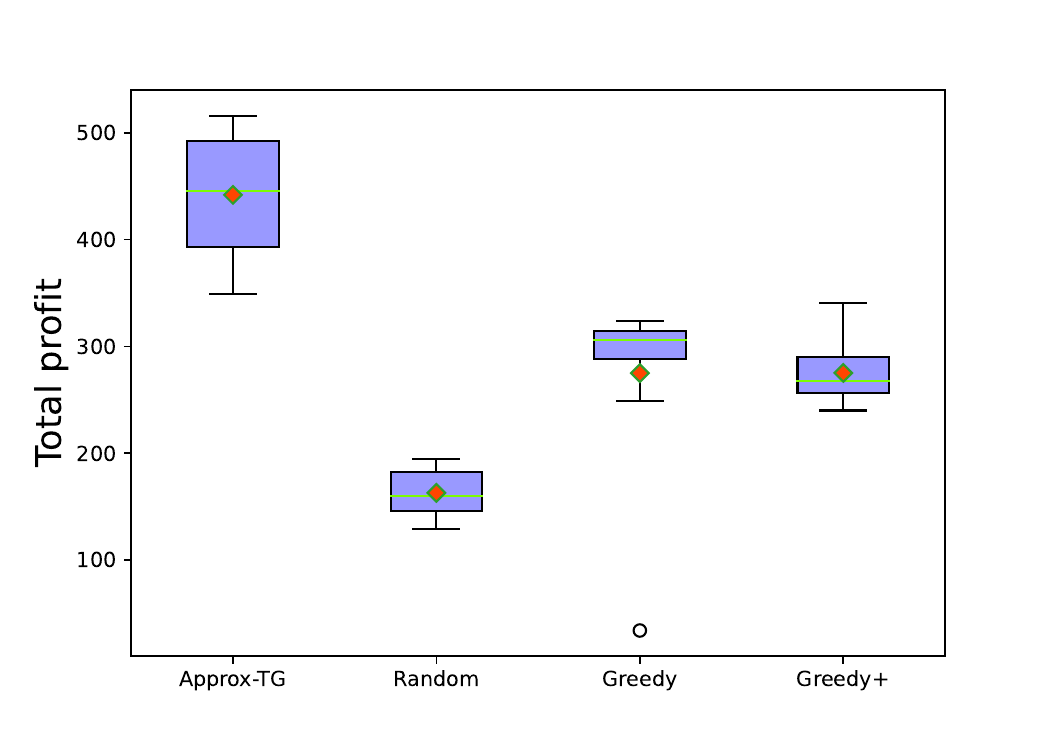}
		\caption{with social compatibility}
		\label{Fig:2}
	\end{figure}

	In this section, we report the performance of four algorithms under the basic setting. Figure \ref{F1} shows the statistics of 10-times-running without considering social compatibility, Figure \ref{Fig:2} shows the statistics of 10-times-running subject to the constraint of social compatibility.
	
	Table \ref{Tab:1} lists the mean total profit of 10 times and the improvement of other three algorithms over random policy.
	\begin{table}[!h]
		\caption{Profit comparison of four algorithms: mean(improvement against random policy)}\label{Tab:1}
		\begin{center}
			\begin{tabular}{|c|c|c|c|c|}
				\hline Setting & Random& Greedy&Greedy+&Approx-TG \\
				\hline without social compatibility & 236.9 & 331.5(40\%) & 320.4(35\%)& 516.3(118\%) \\
				\hline with social compatibility & 162.9 & 275.0(69\%) & 275.2(69\%)& 441.9(171\%) \\
				\hline
			\end{tabular}
		\end{center}
	\end{table}

Our algorithm outperforms three benchmark solutions under all settings.  
By treating \textsf{Random} as a baseline, \textsf{Greedy}, \textsf{Greedy+} and our algorithm increase the total profit by $40\%$, $35\%$ and $118\%$, respectively, without considering social compatibility. If we consider the constraint of social compatibility,  \textsf{Greedy}, \textsf{Greedy+} and our algorithm increase the total profit by $69\%$, $69\%$ and $171\%$, respectively. The absolute profits achieved by all algorithms decrease as we consider the constraint of social compatibility. This is because considering social compatibility adds additional constraints for finding a feasible solution.

	\begin{table}[!h]
		\caption{Profit comparison of four algorithms: mean(improvement against random policy) under different scenarios}\label{Tab:2}
		\begin{center}
			\begin{tabular}{|c|c|c|c|c|c|}
				\hline index &scenario & Random& Greedy&Greedy+&Approx-TG \\
				\hline \textcircled{1}&(-,10,20,100) & 236.9 & 331.5(40\%) & 320.4(35\%)& 516.3(118\%) \\
				\hline \textcircled{2}&(-,10,20,50) & 96.0  & 165.5 (72\%) & 150.6(57\%)& 230.3 (140\%) \\
				\hline \textcircled{3}&(-,10,20,200) & 464.5 & 668.5 (43\%) & 614.8(32\%)& 973.1(109\%) \\
				\hline \textcircled{4}&(-,10,10,100) & 215.3 & 308.0(43\%) & 286.8(33\%)& 362.3(68\%) \\
				\hline \textcircled{5}&(-,10,50,100) & 236.7 & 348.5(47\%) & 335.9(42\%)& 615.4(160\%) \\
				\hline \textcircled{6}&(-,20,20,100) & 204.1 & 271.3(32\%) & 252.5(23\%)& 337.3(65\%) \\
				\hline
			\end{tabular}
		\end{center}
\label{tab:a}
	\end{table}

	\begin{table}[!htb]
		\caption{Profit comparison of four algorithms: mean(improvement against random policy) under different network scenarios}\label{Tab:3}
		\begin{center}
			\begin{tabular}{|c|c|c|c|c|c|}
				\hline index &scenario & Random& Greedy&Greedy+&Approx-TG \\
				\hline \textcircled{7}&(1000,10,20,100) & 162.9 & 275.0(69\%) & 275.2(69\%)& 441.9(171\%) \\
				\hline \textcircled{8}&(2000,10,20,100) & 186.8 & 353.0(89\%) & 313.0(68\%)& 474.5(154\%) \\
				\hline \textcircled{9}&(3000,10,20,100) & 210.2 & 385.0(83\%) & 337.6(61\%)& 479.8(128\%) \\
				\hline
			\end{tabular}
		\end{center}
\label{tab:b}
	\end{table}
\subsection{More Results on Robustness Check}
	We conduct additional experiments to validate the robustness of our algorithm. We denote the scenario that involves $c_1$ edges, $c_2$ skills, $c_3$ tasks and $c_4$ individuals as $\left(c_1, c_2, c_3, c_4\right)$, and denote with $\left(-, c_2, c_3, c_4\right)$ the same scenario without considering the social compatibility. We vary the number of edges, the number of  skills, the number of tasks and the number of individuals and report the mean total profit of 10-times-running in Tables \ref{tab:a} and \ref{tab:b}.
	
\textsf{Random} performs the worst in all settings, and our algorithm achieves the largest improvement over \textsf{Random}. Table  \ref{tab:a} shows the results without social compatibility. We report the results under the baseline setting $(-,10,20,100)$ in \textcircled{1}. We vary the  number of individuals  from $50$ to $200$ and report the results in  \textcircled{2} and \textcircled{3} respectively. As the same task can be performed by an arbitrary number of teams, a larger pool of individuals leads to a higher profit. 
Our algorithm achieves a profit more than twice that of \textsf{Random}, regardless of the number of individuals. We vary the number of tasks  in  \textcircled{4} and \textcircled{5}. We find that the number of tasks has little impact on the performance of three benchmarks. Both \textsf{Greedy} and \textsf{Greedy+} tend to select tasks with higher profits. \textsf{Approx-TG} achieves an improvement of $160\%$ when the number of tasks is large. We increase the number of skills to $20$ in  \textcircled{6}. 
While the performance of all algorithms decline, our algorithm still performs the best.
	
Considering the constraint of social compatibility, we conduct experiments  under different number of edges and report our results in Table  \ref{tab:b}. We set the number of edges to $2000$ and $3000$ in \textcircled{8} and \textcircled{9}, respectively. As we add more edges to the network, it is easier to form a feasible team for a given task, this improves the profit of all algorithms. Our algorithm still performs the best, i.e., it achieves a profit more than twice that achieved by \textsf{Random}.

\section{Extensions}
\label{sec:extend}

\subsection{Incorporation of the Capacity Constraint of Each Task}
So far, we have assumed that each task can be performed an unlimited number of times. However, this may not always hold in practice. For example, puzzle assembly can only be performed once. To this end, we add a group of additional constraints to the original problem: $\sum_{C_{ti}\in \mathcal{C}_t } x_{ti} \leq g_t, \forall t\in \mathcal{T}$, where $g_t$ denotes the capacity of task $t\in \mathcal{T}$; that is, each task $ t\in \mathcal{T}$ can be performed up to $g_t$ times.
We formally define this extension in \textbf{P.2}.
\begin{center}
\framebox[0.78\textwidth][c]{
\enspace
\begin{minipage}[t]{0.45\textwidth}
\small
\textbf{P.2:}
\emph{Maximize $\sum_{C_{ti}\in \mathcal{C}}(x_{ti}\cdot \lambda_t)$}\\
\textbf{subject to:}
\begin{equation*}
\begin{cases}
 \sum_{C_{ti}\in \mathcal{C}: C_{ti}\ni u} x_{ti} \leq 1, \forall u\in \mathcal{V}\\
   \sum_{C_{ti}\in \mathcal{C}_t } x_{ti} \leq g_t, \forall t\in \mathcal{T}\\
 x_{ti} \in \{0,1\}, \forall C_{ti}\in  \mathcal{C}.

\end{cases}
\end{equation*}
\end{minipage}
}
\end{center}
\vspace{0.1in}

Similar to the  LP-Based algorithm developed in Section \ref{sec:LP}, we propose a LP-Based algorithm for \textbf{P.2}.

\textbf{LP Relaxation} The primal LP of \textbf{P.2} can be formulated as follows.
\begin{center}
\framebox[0.78\textwidth][c]{
\enspace
\begin{minipage}[t]{0.45\textwidth}
\small
\textbf{Primal LP of P.2:}
\emph{Maximize $\sum_{C_{ti}\in \mathcal{C}}(x_{ti}\cdot \lambda_t)$}\\
\textbf{subject to:}
\begin{equation*}
\begin{cases}
 \sum_{C_{ti}\in \mathcal{C}: C_{ti}\ni u} x_{ti} \leq 1, \forall u\in \mathcal{V}\\
  \sum_{C_{ti}\in \mathcal{C}_t } x_{ti} \leq g_t, \forall t\in \mathcal{T}\\
 0 \leq x_{ti}, \forall C_{ti} \in \mathcal{C}.

\end{cases}
\end{equation*}
\end{minipage}
}
\end{center}
\vspace{0.1in}
The dual to the above primal LP assigns a
price $y(u)$ to each node $u \in V$ and a price $p(t)$ to each task $t\in \mathcal{T}$.
\begin{center}
\framebox[0.78\textwidth][c]{
\enspace
\begin{minipage}[t]{0.55\textwidth}
\small
\textbf{Dual LP of P.2:} \emph{Minimize $\sum_{u\in \mathcal{V}} y(u) + \sum_{t\in \mathcal{T}} (p(t)\cdot g_t) $}\\
\textbf{subject to:}
\begin{equation*}
\begin{cases}
 \sum_{u\in C_{ti}} y(u) +p(t) \geq \lambda_t, \forall C_{ti} \in \mathcal{C}\\
y(u) \geq 0, \forall u \in \mathcal{V}; p(t) \geq 0, \forall t \in \mathcal{T}.

\end{cases}
\end{equation*}
\end{minipage}
}
\end{center}
\vspace{0.1in}

Similar to the solution for \textbf{P.1}, we run the ellipsoid algorithm on the dual LP
using algorithm $\mathcal{A}$, an approximation algorithm for
\textsc{MincostTeamSelection}, as the approximate separation
oracle. More precisely, let $S (L)$ denote the set of $y\in \mathbb{R}_+^\mathcal{V}$ satisfying that
\[\sum_{u\in \mathcal{V}} y(u) + \sum_{t\in \mathcal{T}} (p(t)\cdot g_t) \leq L,\]
\[ \sum_{u\in C_{ti}} y(u) +p(t) \geq \lambda_t, \forall C_{ti} \in \mathcal{C}.\]
We adopt binary search to find the smallest value of $L$ for
which $S (L)$ is nonempty. The separation oracle works as
follows: First, it checks the inequality $\sum_{u\in \mathcal{V}} y(u) + \sum_{t\in \mathcal{T}} (p(t)\cdot g_t) \leq L$. Next, it runs the algorithm $\mathcal{A}$ on each task $t\in \mathcal{T}$ and selects a group $C_{ti} \in \mathcal{C}_t, \forall t\in \mathcal{T}$, using $y(u)$ as the price function. If for all $C_{ti}$, the cost of $C_{ti}$ is larger than $\lambda_t-p(t)$, then $y\in S(L)$. If  there exists some $C_{ti}$ whose cost is less than $\lambda_t-p(t)$, then $y\notin S(L)$ and $C_{ti}$ gives us a separating hyperplane. Based on similar analysis in Section \ref{sec:LP}, we have the following theorem.

\begin{theorem}
If there is a polynomial $\mu$-approximation algorithm for \textsc{MincostTeamSelection}, then there exists a polynomial $\mu$-approximation algorithm for \textbf{Primal LP of P.2}.
\end{theorem}

\textbf{LP Rounding} 
We present a deterministic rounding algorithm (Algorithm \ref{alg:LPP6}). Given a feasible solution $x^*$ of \textbf{Primal LP of P.2}, Algorithm \ref{alg:LPP6} takes $\mathcal{C}^{I}$, a subset of $\mathcal{C}(x^*)$, as input, and outputs a group of teams from $\mathcal{C}^{I}$ such that (1) each individual participates in at most one team and (2) the same task $t$ is performed by at most $g_t$ teams for each task $t\in \mathcal{T}$. We next provide a summary of Algorithm \ref{alg:LPP6}:


Initially, let $\mathcal{C}^{DR}=\emptyset, z=x^*$.

\emph{Step 1:} Select the team with the highest profit from $\mathcal{C}^{I}$ (e.g., $C_{ti}$).


\emph{Step 2:} Let $\mathcal{C}^{I}_t =\mathcal{C}_t\cap  \mathcal{C}^{I}$ denote the set of all teams  in $\mathcal{C}^{I}$ that is assigned to task $t$. Reduce the value of $z_{tj}$ for some $C_{tj} \in \mathcal{C}^{I}_t\setminus \{C_{ti}\}$ to some non-negative value such that
\begin{eqnarray}
\label{cond:!}
\sum_{C_{tj} \in \mathcal{C}^{I}_t\setminus \{C_{ti}\}} z_{tj}\mbox{ is reduced by }\min\{\sum_{C_{tj} \in \mathcal{C}^{I}_t\setminus \{C_{ti}\}} z_{tj}, 1-z_{ti}\}.
 \end{eqnarray}

 This can be done in an arbitrary way. For example, one can select an arbitrary team, say $C_{tq}$, from $\mathcal{C}^{I}_t\setminus \{C_{ti}\}$, reduce $z_{tq}$ to its smallest non-negative value  (zero, if necessary) such that the cumulative  amount of reduction does not exceed $\min\{\sum_{C_{tj} \in \mathcal{C}^{I}_t\setminus \{C_{ti}\}} z_{tj}, 1-z_{ti}\}$. Then we select another team from  $\mathcal{C}^{I}_t\setminus \{C_{ti}\}$ and reduce its fractional value in the same manner. This process iterates until condition (\ref{cond:!}) is satisfied; that is, we terminate this process once  the cumulative  amount of reduction reaches $\min\{\sum_{C_{tj} \in \mathcal{C}^{I}_t\setminus \{C_{ti}\}} z_{tj}, 1-z_{ti}\}$. 

\emph{Step 3:} Recall that $\mathcal{N}(C_{ti})$ denotes the set of all adjacent teams of $C_{ti}$ from  $\mathcal{C}^{I}\subseteq \mathcal{C}(x^*)$. Remove  $\mathcal{N}(C_{ti}) \cup \{C_{ti}\}$ and $\{ C_{tj} \in \mathcal{C}^{I}_t\} \mid  z_{tj}=0\}$ from $\mathcal{C}^{I}$. It will become clear later that this step ensures that no individual participates in multiple tasks and meanwhile, each task is assigned to at most $g_t$ teams.


\emph{Step 4:} Go to Step 1 unless $\mathcal{C}^{I}$ becomes empty. Output $\mathcal{C}^{DR}$.

\begin{algorithm}[hptb]
\caption{Deterministic Rounding}
\label{alg:LPP6}
\textbf{Input}: $x^*, \mathcal{C}^{I}\subseteq \mathcal{C}(x^*)$.
\begin{algorithmic}[1]
\STATE $\mathcal{C}^{DR}=\emptyset, z=x^*$.
\WHILE {$\mathcal{C}^{I} \neq \emptyset$}
\STATE Select the team  that has the highest profit from $\mathcal{C}^{I}$ (e.g., $C_{ti}$).
\STATE $\mathcal{C}^{DR} = \mathcal{C}^{DR} \cup \{C_{ti}\}$.
\STATE Reduce  $z_{tj}$ for some $C_{tj} \in \mathcal{C}^{I}_t\setminus \{C_{ti}\}$ to satisfy condition (\ref{cond:!}).
\STATE Remove  $\mathcal{N}(C_{ti}) \cup \{C_{ti}\}$ and  $\{ C_{tj} \in \mathcal{C}^{I}_t\ \mid  z_{tj}=0\}$ from $\mathcal{C}^{I}$.
\ENDWHILE
\STATE Output $\mathcal{C}^{DR}$.
\end{algorithmic}
\end{algorithm}

Let $\mathcal{C}^{DR}$ denote the output of Algorithm \ref{alg:LPP6}, we first show that  $\mathcal{C}^{DR}$ is a feasible solution to \textbf{P.2}.
\begin{lemma}
\label{the:feasible}
Let $\mathcal{C}^{DR}$ denote the set of groups returned from the deterministic rounding (Algorithm \ref{alg:LPP6}). $\mathcal{C}^{DR}$ is a feasible solution to \textbf{P.2}.
\end{lemma}
\emph{Proof:} First, by the design of Algorithm \ref{alg:LPP6}, once a team is selected, we remove all its adjacent teams from consideration. Hence, in the final solution  $\mathcal{C}^{DR}$, each individual participates in at most one team. We next show that $\mathcal{C}^{DR}$  satisfies the capacity constraint of each task. To prove this, we show that for each task $t\in \mathcal{T}$,  $\mathcal{C}^{I}_t$ becomes empty after adding at most $g_t$ number of teams from $\mathcal{C}_t$ to $\mathcal{C}^{DR}$, 
 indicating that no more teams from  $\mathcal{C}_t$ will be added to the solution. Assume by contradiction that after selecting a group $\mathcal{C}'$ of $g_t$ teams from  $\mathcal{C}^{I}_t$,
\begin{eqnarray}\label{eq:frisco}
\sum_{C_{ti}\in \mathcal{C}^{I}_t\setminus \mathcal{C}'}z_{ti}>0
 \end{eqnarray}
Recall that after selecting a team $C_{ti}$, we reduce the value of $\sum_{C_{tj} \in \mathcal{C}^{I}_t\setminus \{C_{ti}\}} z_{tj}$ by an amount of $\min\{\sum_{C_{tj} \in \mathcal{C}^{I}_t\setminus \{C_{ti}\}} z_{tj}, 1-z_{ti}\}$. Because of (\ref{eq:frisco}), we conclude that the cumulative  amount of reduction in $z$ due to the selection of $\mathcal{C}'$ is exactly $\sum_{C_{ti}\in \mathcal{C}'}(1-x^*_{ti})$. It follows that
\begin{eqnarray}
\sum_{C_{ti}\in \mathcal{C}_t}x^*_{ti}&\geq& \sum_{C_{ti}\in \mathcal{C}'}1+\sum_{C_{ti}\in \mathcal{C}^{I}_t\setminus \mathcal{C}'}z_{ti} \\
&\geq& g_t + \sum_{C_{ti}\in \mathcal{C}^{I}_t\setminus \mathcal{C}'}z_{ti} \\
&>& g_t,
\end{eqnarray}
where the second inequality is due to the assumption that $|\mathcal{C}'|=g_t$ and the third inequality is due to (\ref{eq:frisco}). This contradicts to the assumption that $x^*$ is a feasible solution to \textbf{Primal LP of P.2}; that is, $x^*$ violates the second set of constraints listed in \textbf{Primal LP of P.2}.
$\Box$

 We next show that the profit of $\mathcal{C}^{DR}$ is at least $1/(\rho+1)$ of the one obtained from the fractional solution $x^*$. That is,
\begin{lemma}
\label{the:constant1}
Given a feasible solution $x^*$ of \textbf{Primal LP of P.2}, a set of input teams $\mathcal{C}^{I}\subseteq \mathcal{C}(x^*)$, let $\mathcal{C}^{DR}$ denote the set of groups returned from the deterministic rounding (Algorithm \ref{alg:LPP6}), $\sum_{C_{ti} \in \mathcal{C}^{DR}} \lambda_t \geq \sum_{C_{ti} \in \mathcal{C}(x^*)}(x^*_{ti}\cdot \lambda_t)/(\rho+1)$, where $\rho=\max_{C_{ti}\in \mathcal{C}^{I}}|C_{ti}|$.
\end{lemma}
\emph{Proof:} Consider an arbitrary team from $\mathcal{C}^{DR}$ (e.g., $C_{ti}$). According to the design of Algorithm \ref{alg:LPP6}, after selecting $C_{ti}$, we perform the following two operations that may cause profit loss: (a) remove $\mathcal{N}(C_{ti}) \cup \{C_{ti}\}$, and (b) reduce the value of $\sum_{C_{tj} \in \mathcal{C}^{I}_t\setminus \{C_{ti}\}} z_{tj}$ by an amount of $\min\{\sum_{C_{tj} \in \mathcal{C}^{I}_t\setminus \{C_{ti}\}} z_{tj}, 1-z_{ti}\}$. We next bound the profit loss due to these two operations separately.

  First, by the design of Algorithm \ref{alg:LPP6}, $C_{ti}$ has the highest profit among $\mathcal{N}(C_{ti}) \cup \{C_{ti}\}$. Following the same proof of (\ref{eq:aaai}), we  can bound the amount of profit loss $z_{ti}\cdot \lambda_t + \sum_{C_{lj}\in \mathcal{N}(C_{ti})} (z_{li}\cdot \lambda_l)$ due to the removal of  $\mathcal{N}(C_{ti}) \cup \{C_{ti}\}$ as follows:
\begin{equation}
\label{eq:1.1}
z_{ti}\cdot \lambda_t + \sum_{C_{lj}\in \mathcal{N}(C_{ti})} (z_{li}\cdot \lambda_l) \leq \rho \cdot \lambda_t.
 \end{equation}

 On the other hand, because all teams in $\mathcal{C}^{I}_t$ have equal profit $\lambda_t$, the amount of the reduced profit due to  operation (b) is $\lambda_t\times \min\{\sum_{C_{tj} \in \mathcal{C}^{I}_t\setminus \{C_{ti}\}} z_{tj}, 1-z_{ti}\}$. We next show that this value is at most $\lambda_t$.
 \begin{equation}
\label{eq:1.2}
\lambda_t\times \min\{\sum_{C_{tj} \in \mathcal{C}^{I}_t\setminus \{C_{ti}\}} z_{tj}, 1-z_{ti}\} \leq \lambda_t\times (1-z_{ti}) \leq \lambda_t,
\end{equation}
 where the third inequality is due to $1-z_{ti}\leq 1$.

Eqs. (\ref{eq:1.1}) and (\ref{eq:1.2}) together imply that
\[(1+\rho)\lambda_t \geq (z_{ti}\cdot \lambda_t + \sum_{C_{lj}\in \mathcal{N}(C_{ti})} (z_{lj}\cdot \lambda_l))+\lambda_t\cdot \min\{\sum_{C_{tj} \in \mathcal{C}^{I}_t\setminus \{C_{ti}\}} z_{tj}, 1-z_{ti}\}.\]

It follows that

 \begin{equation}
\label{eq:mainevent}\lambda_t \geq \frac{1}{1+\rho}\left( (z_{ti}\cdot \lambda_t + \sum_{C_{lj}\in \mathcal{N}(C_{ti})} (z_{lj}\cdot \lambda_l))+\lambda_t\cdot \min\{\sum_{C_{tj} \in \mathcal{C}^{I}_t\setminus \{C_{ti}\}} z_{tj}, 1-z_{ti}\}\right).
\end{equation}

Note that $\lambda_t$ represents the profit of $C_{ti}$ and $(z_{ti}\cdot \lambda_t + \sum_{C_{lj}\in \mathcal{N}(C_{ti})} (z_{lj}\cdot \lambda_l))+\lambda_t\cdot \min\{\sum_{C_{tj} \in \mathcal{C}^{I}_t\setminus \{C_{ti}\}} z_{tj}, 1-z_{ti}\}$ represents the amount of reduced profit due to the selection of $C_{ti}$. Hence,  (\ref{eq:mainevent}) indicates that we retain at least $1/(1+\rho)$ fraction of the original profit after selecting $C_{ti}$. Summing up (\ref{eq:mainevent}) over all teams from $\mathcal{C}^{DR}$ gives  $\sum_{C_{ti} \in \mathcal{C}^{DR}} \lambda_t \geq \sum_{C_{ti} \in \mathcal{C}^{H}}(x^*_{ti}\cdot \lambda_t)/(\rho+1)$. $\Box$

\paragraph{Algorithm Design and Performance Analysis}
\textsf{Approx-TG}  can be naturally adapted to handle this generalization by replacing its
LP rounding method with Algorithm \ref{alg:LPP6}. In analogy to Lemmas \ref{lem:1} and \ref{lem:2}, we present two lemmas to prove the performance bounds of the first and the second candidate solutions, respectively.
\begin{lemma}
\label{lem:1-1}
Assume $x^*$ is a $\mu$-approximate solution of \textbf{Primal LP of P.2} that is found by the ellipsoid method, Algorithm \ref{alg:LPP1} (whose LP rounding method is replaced with Algorithm \ref{alg:LPP6})  achieves an approximation ratio of $\mu/(\Delta+1)$  for \textbf{P.2}.
\end{lemma}
\emph{Proof:} By Lemma \ref{the:constant1}, our deterministic rounding technique (Algorithm \ref{alg:LPP6}), taking $\mathcal{C}(x^*)$ as input, returns a grouping that achieves a profit of at least $\frac{1}{\Delta+1}  \sum_{C_{ti}\in \mathcal{C}(x^*)} (x^*_{ti}\cdot \lambda_t)$, where  $\Delta$ is the size of the largest possible team in $\mathcal{C}(x^*)$. By the assumption that  $x^*$ is a $\mu$-approximate solution of \textbf{Primal LP of P.2}, we have $\sum_{C_{ti}\in \mathcal{C}(x^*)} (x^*_{ti}\cdot \lambda_t)\geq \mu\cdot OPT$. It follows that Algorithm \ref{alg:LPP1} achieves a profit of at least
\[\frac{1}{\Delta+1}  \sum_{C_{ti}\in \mathcal{C}(x^*)} (x^*_{ti}\cdot \lambda_t)\geq \frac{\mu}{\Delta+1}\cdot OPT.\]
 $\Box$

\begin{lemma}
\label{lem:2-1}
Assume $x^*$ is a $\mu$-approximate solution of \textbf{Primal LP of P.2} that is found by the ellipsoid method, Algorithm \ref{alg:LPP2} (whose LP rounding method is replaced with Algorithm \ref{alg:LPP6})  achieves an approximation ratio  of $\mu  /2(\sqrt{m}+1)$  for \textbf{P.2}.
\end{lemma}
\emph{Proof:}
To prove this lemma, it suffices to show that $\max\{\sum_{C_{ti}\in \widetilde{\mathcal{C}}} \lambda_t, \lambda_{t_{\max}}\}\geq \frac{1}{\sqrt{m}+1}\cdot \frac{\mu }{2}\cdot OPT$.

We first bound the gap between the profit gained from $\widetilde{\mathcal{C}}$  and $\sum_{C_{ti}\in  \mathcal{C}(x^*)^1}(x^*_{ti}\cdot \lambda_t)$. By Lemma \ref{the:constant1}, we have
\begin{equation}
\label{eq:1-1}\sum_{C_{ti}\in \widetilde{\mathcal{C}}} \lambda_t \geq \frac{1}{\rho(\widetilde{\mathcal{C}})+1}\cdot \sum_{C_{ti} \in \mathcal{C}(x^*)^1}(x^*_{ti}\cdot \lambda_t ) \geq \frac{1}{\sqrt{m}+1}\cdot \sum_{C_{ti}\in \mathcal{C}(x^*)^1}(x^*_{ti}\cdot \lambda_t ) \end{equation}
 where the second inequality is due to the assumption that $\rho(\widetilde{\mathcal{C}}) \leq \sqrt{m}$ holds for all teams from $\mathcal{C}(x^*)^1$.

Adopting the same argument used in the proof of Lemma \ref{lem:2}, we can bound the gap between the profit gained from $C_{t_{\max}}$ and $\sum_{C_{ti}\in  \mathcal{C}(x^*)^2}(x^*_{ti}\cdot \lambda_t)$ as follows:
\begin{eqnarray}
\label{eq:211-1}
\lambda_{t_{\max}}\geq \frac{1}{\sqrt{m}} \cdot \sum_{C_{ti} \in \mathcal{C}(x^*)^2}(x^*_{ti}\cdot \lambda_t).
\end{eqnarray}

By the assumption that $x^*$ is a $\mu$-approximate solution of \textbf{Primal LP of P.2}, we have
\begin{equation}
\label{eq:3-1}\sum_{C_{ti} \in \mathcal{C}(x^*)}(x^*_{ti}\cdot \lambda_t) \geq \mu \cdot OPT \Rightarrow \sum_{C_{ti} \in \mathcal{C}(x^*)^1}(x^*_{ti} \cdot \lambda_t) + \sum_{C_{ti} \in \mathcal{C}(x^*)^2}(x^*_{ti} \cdot \lambda_t) \geq \mu \cdot OPT.\end{equation}

Now we are ready to prove this theorem.
\begin{eqnarray}
&&\max\{\sum_{C_{ti}\in \widetilde{\mathcal{C}}} \lambda_t, \lambda_{t_{\max}}\}\geq \frac{\sum_{C_{ti}\in \widetilde{\mathcal{C}}} \lambda_t+ \lambda_{t_{\max}}}{2}\\
&&\geq  \frac{ \frac{1}{\sqrt{m}+1}\cdot  \sum_{C_{ti} \in \mathcal{C}(x^*)^1}(x^*_{ti} \cdot \lambda_t) + \frac{1}{\sqrt{m}} \cdot \sum_{C_{ti} \in \mathcal{C}(x^*)^2}(x^*_{ti}\cdot \lambda_t) }{2}\\
&&\geq \frac{1}{\sqrt{m}+1}\cdot \frac{ \sum_{C_{ti} \in \mathcal{C}(x^*)^1}(x^*_{ti} \cdot \lambda_t) + \sum_{C_{ti} \in \mathcal{C}(x^*)^2}(x^*_{ti}\cdot \lambda_t) }{2}\\
&&\geq \frac{1}{\sqrt{m}+1}\cdot \frac{\mu }{2}\cdot OPT
\end{eqnarray}
where the second inequality is due to (\ref{eq:1-1}) and (\ref{eq:211-1}), and the last inequality is due to  (\ref{eq:3-1}). $\Box$

Lemma \ref{lem:1-1} and Lemma \ref{lem:2-1} together imply  the following theorem.
\begin{theorem}
Approx-TG achieves an  approximation ratio of $\max\{\mu/(\Delta+1), \mu  /2(\sqrt{m}+1)\}$ for \textbf{P.2}.
\end{theorem}

\subsection{Incorporation of Heterogenous Load Limits}
Our basic model assumes that each individual can only participate in \emph{one} task. For the general case when each individual $u$ can participate in up to $f_u$ number of tasks,  one naive approach is to  simply create $f_u$ copies of $u$ with identical skill set for each $u$. It turns out we can still apply Approx-TG to this expanded set to achieve an approximation ratio of $\max\{\mu/\Delta, \mu  /2\sqrt{m}\}$. However, this is not a polynomial time algorithm if $f_u$ is exponential in the size of input. We next present a polynomial time approximation algorithm based on LP relaxation.

\begin{center}
\framebox[0.78\textwidth][c]{
\enspace
\begin{minipage}[t]{0.45\textwidth}
\small
\textbf{P.3:}
\emph{Maximize $\sum_{C_{ti}\in \mathcal{C}}(x_{ti}\cdot \lambda_t)$}\\
\textbf{subject to:}
\begin{equation*}
\begin{cases}
 \sum_{C_{ti}\in \mathcal{C}: C_{ti}\ni u} x_{ti} \leq f_u, \forall u\in \mathcal{V}\\
 x_{ti} \in \{0,1\}, \forall C_{ti}\in  \mathcal{C}.

\end{cases}
\end{equation*}
\end{minipage}
}
\end{center}
\vspace{0.1in}

\textbf{LP Relaxation} The primal LP of \textbf{P.3} can be formulated as follows.
\begin{center}
\framebox[0.78\textwidth][c]{
\enspace
\begin{minipage}[t]{0.45\textwidth}
\small
\textbf{Primal LP of P.3:}
\emph{Maximize $\sum_{C_{ti}\in \mathcal{C}}(x_{ti}\cdot \lambda_t)$}\\
\textbf{subject to:}
\begin{equation*}
\begin{cases}
 \sum_{C_{ti}\in \mathcal{C}: C_{ti}\ni u} x_{ti} \leq f_u, \forall u\in \mathcal{V}\\
 0 \leq x_{ti} , \forall C_{ti}\in  \mathcal{C}.

\end{cases}
\end{equation*}
\end{minipage}
}
\end{center}
\vspace{0.1in}

In the dual problem, we assign a
price $y(u)$ to each node $u \in \mathcal{V}$:
\begin{center}
\framebox[0.78\textwidth][c]{
\enspace
\begin{minipage}[t]{0.45\textwidth}
\small
\textbf{Dual LP of P.3:} \emph{Minimize $\sum_{u\in \mathcal{V}} f_u\cdot y(u)$}\\
\textbf{subject to:}
\begin{equation*}
\begin{cases}
 \sum_{u\in C_{ti}} y(u) \geq \lambda_t, \forall C_{ti}\in \mathcal{C}\\
y(u) \geq 0, \forall u \in \mathcal{V}.

\end{cases}
\end{equation*}
\end{minipage}
}
\end{center}
\vspace{0.1in}

We can still adopt the ellipsoid method for exponential-sized LP with an (approximate) separation
oracle to solve \textbf{Dual LP of P.3} to obtain a fractional solution $x^*$.

\begin{theorem}
If there is a polynomial $\mu$-approximation algorithm for \textsc{MincostTeamSelection}, then there exists a polynomial $\mu$-approximation algorithm for \textbf{Primal LP of P.3}.
\end{theorem}

\textbf{LP Rounding} Our randomized rounding (Algorithm~\ref{alg:LPP}) consists of two stages: a \emph{initial rounding stage}
and a \emph{conflict resolution stage}. In the initial rounding stage, we covert $x^*$ to a group of teams that might not be feasible; then in the second stage, we remove some teams to obtain a feasible solution. We next explain each stage in detail. Given a feasible solution $x^*$ of \textbf{Primal LP of P.3}, Algorithm~\ref{alg:LPP} takes $\mathcal{C}^{I}$, a subset of $\mathcal{C}(x^*)$, as input.
\begin{enumerate}
\item For each team  $C_{ti}\in \mathcal{C}^{I}$, add $C_{ti}$ to $\mathcal{C}^{RR}$ with probability $\frac{x_{ti}^*}{2\rho}$, where $\rho$ is the size of the largest team in $\mathcal{C}^{I}$. We say $C_{ti}$ survives in the first stage if $C_{ti}$ has been added to $\mathcal{C}^{RR}$.
\item Note that $\mathcal{C}^{RR}$ might violate the constraint of load limits. This can be resolved as follows: For each team $C_{ti}\in \mathcal{C}^{RR}$, keep $C_{ti}$ in $\mathcal{C}^{RR}$ if and only if for all $u\in C_{ti}$,  $\sum_{C_{lj}\in \mathcal{C}^{RR}: C_{lj}\ni u} 1 \leq f_u$. We say $C_{ti}$ survives in the second stage if $C_{ti}$ has been kept in $\mathcal{C}^{RR}$. Return   $\mathcal{C}^{RR}$ as the output. 
\end{enumerate}
\begin{algorithm}[hptb]
\caption{Randomized Rounding}
\label{alg:LPP}
\textbf{Input}: $x^*, \mathcal{C}^{I}\subseteq \mathcal{C}(x^*)$.
\begin{algorithmic}[1]
\STATE $\mathcal{C}^{RR}=\emptyset$.
\FOR {each team $C_{ti}$ in $\mathcal{C}^{I}$}
\STATE Add $C_{ti}$ to $\mathcal{C}^{RR}$ with probability $\frac{x_{ti}^*}{2\rho}$.
\ENDFOR
\FOR {each $C_{ti}\in \mathcal{C}^{RR}$}
\STATE Remove $C_{ti}$ from $\mathcal{C}^{RR}$ if for some $u\in C_{ti}$,  $\sum_{C_{lj}\in \mathcal{C}^{RR}: C_{lj}\ni u} 1 > f_u$.
\ENDFOR
\STATE Return   $\mathcal{C}^{RR}$.
\end{algorithmic}
\end{algorithm}

\begin{lemma}
\label{lem:LP1}
Given any feasible solution $x^*$ of \textbf{Primal LP of P.3} and input teams $\mathcal{C}^{I}\subseteq \mathcal{C}(x^*)$, for each team  $C_{ti}\in \mathcal{C}^{I}$,  $C_{ti}$ survives in the first stage with probability $\frac{x_{ti}^*}{2\rho}$.
\end{lemma}

The above lemma can be directly derived from our algorithm description. Next we use 0/1 random variable $X_{ti}$ to indicate whether  $C_{ti}$ has survived in the first stage, we can immediately have $\mathbf{E}[X_{ti}] = \frac{x_{ti}^*}{2\rho}$.

\begin{lemma}
\label{lem:LP2}
For any team $C_{ti}\in \mathcal{C}^{I}$ that is having survived in the first stage, the probability that $C_{ti}$ still survives in the second stage is at least $\frac{1}{2}$.
\end{lemma}

\emph{Proof:} For each $C_{ti}\in \mathcal{C}^{I}$, let $Y_{ti}$ be a 0/1 random variable representing whether  $C_{ti}$ has survived in the second stage. The event that $C_{ti}$ survives in the first phase but removed in the second stage can be represented as: $Y_{ti} = 0$, under the condition that $X_{ti} = 1$. And the probability of this event is  $\Pr[Y_{ti}=0|X_{ti}=1]$. We note that this event can only happen if  for some $u\in C_{ti}$,
\[\sum_{C_{lj}\in \mathcal{C}^{I}\setminus C_{ti}: C_{lj}\ni u} X_{lj}\geq f_u.\]

By Markov's inequality, the probability of this event can be bounded by
\begin{eqnarray}
\Pr[Y_{ti}=0|X_{ti}=1] &\leq& \sum_{u\in  C_{ti}}\Pr[\sum_{C_{lj}\in \mathcal{C}^{I}\setminus C_{ti}: C_{lj}\ni u} X_{lj}\geq f_u] \\
&\leq&\sum_{u\in  C_{ti}} \frac{\mathbf{E}[\sum_{C_{lj}\in \mathcal{C}^{I}\setminus C_{ti}: C_{lj}\ni u} X_{lj}]}{f_u}.\label{eq:53}
\end{eqnarray}
Based on linearity of expectation and $\mathbf{E}[X_{lj}] = \frac{x_{lj}^*}{2\rho}$, for each $u\in  C_{ti}$, we have
\begin{eqnarray}\mathbf{E}[\sum_{C_{lj}\in \mathcal{C}^{I}\setminus C_{ti}: C_{lj}\ni u} X_{lj}] = \sum_{C_{lj}\in \mathcal{C}^{I}\setminus C_{ti}: C_{lj}\ni u} \mathbf{E}[ X_{lj}] = \sum_{C_{lj}\in \mathcal{C}^{I}\setminus C_{ti}: C_{lj}\ni u} \frac{x_{lj}^*}{2\rho}.\label{eq:waco}\end{eqnarray}

By the first constraint of \textbf{Primal LP of P.3}, we further have
\begin{eqnarray}\sum_{C_{lj}\in \mathcal{C}^{I}\setminus C_{ti}: C_{lj}\ni u} \frac{x_{lj}^*}{2\rho} = \frac{\sum_{C_{lj}\in \mathcal{C}^{I}\setminus C_{ti}: C_{lj}\ni u}  x_{lj}^*}{2\rho} \leq \frac{\sum_{C_{lj}\in \mathcal{C}: C_{lj}\ni u}  x_{lj}^*}{2\rho} \leq \frac{f_u}{2\rho}.\label{eq:tamu} \end{eqnarray}

Hence,
\begin{eqnarray}
\Pr[Y_{ti}=0|X_{ti}=1] &\leq&  \sum_{u\in  C_{ti}} \frac{\mathbf{E}[\sum_{C_{lj}\in \mathcal{C}^{I}\setminus C_{ti}: C_{lj}\ni u} X_{lj}]}{f_u}\\
&=& \sum_{u\in  C_{ti}} (\sum_{C_{lj}\in \mathcal{C}^{I}\setminus C_{ti}: C_{lj}\ni u} \frac{x_{lj}^*}{2\rho})\cdot \frac{1}{f_u}\\
&\leq& \sum_{u\in  C_{ti}} \frac{f_u}{2\rho}\cdot \frac{1}{f_u}\\
&\leq& 1/2
\end{eqnarray}
where the first inequality is due to (\ref{eq:53}), the equality is due to (\ref{eq:waco}), the second inequality is due to (\ref{eq:tamu}), and the last inequality is due to $| C_{ti}|\leq \rho$.

Therefore, the probability that each team that survives in the first stage still survives in the second stage is at least $1-\frac{1}{2}=\frac{1}{2}$. $\Box$

Lemma \ref{lem:LP1} and Lemma \ref{lem:LP2} together imply that for each  $C_{ti}\in \mathcal{C}^{I}$, it survives in both stages with probability $x^*_{ti}/(4 \rho)$, hence, the following theorem follows.

\begin{lemma}
\label{the:constant2}
Given a feasible solution $x^*$ of \textbf{Primal LP of P.3}, a set of input teams $\mathcal{C}^{I}\subseteq \mathcal{C}(x^*)$, let $\mathcal{C}^{RR}$ denote the group of teams returned from the randomized rounding (Algorithm \ref{alg:LPP}), $\sum_{C_{ti} \in \mathcal{C}^{RR}} \lambda_t \geq \frac{1}{4\rho}\cdot \sum_{C_{ti} \in \mathcal{C}^{I}}(x^*_{ti}\cdot \lambda_t)$, where $\rho=\max_{C_{ti}\in \mathcal{C}^{I}}|C_{ti}|$.
\end{lemma}

\paragraph{Algorithm Design and Performance Analysis}
We still use \textsf{Approx-TG}  to handle this extension. However, we make two crucial modifications to its original version developed in Section \ref{sec:444} as follows. First, we replace its LP rounding method with Algorithm \ref{alg:LPP}. Second, we modify the second candidate solution (Algorithm \ref{alg:LPP2}) such that we adopt a different criterion to partition each $\mathcal{C}(x^*)_t$. In particular, for every task $t\in \mathcal{T}$, we partition $\mathcal{C}(x^*)_t$ into two disjoint subsets $\mathcal{C}(x^*)^1_t$ and $\mathcal{C}(x^*)^2_t$ such that: $\forall C \in \mathcal{C}(x^*)^1_t: |C|\leq \sqrt{f_{\max}m}$ and $\forall C \in \mathcal{C}(x^*)^2_t: |C| > \sqrt{f_{\max}m}$, where $f_{\max}=\max_{ u\in \mathcal{V}} f_u$ represents the largest number of tasks an individual can participate in. The rest of the algorithm is identical to its original version. 

 In analogy to Lemmas \ref{lem:1} and \ref{lem:2}, we present two lemmas to prove the performance bounds of the first and the second candidate solutions, respectively.
\begin{lemma}
\label{lem:1-2}
Assume $x^*$ is a $\mu$-approximate solution of \textbf{Primal LP of P.3} that is found by the ellipsoid method, Algorithm \ref{alg:LPP1} (whose LP rounding method is replaced with Algorithm \ref{alg:LPP})  achieves an approximation ratio of $\mu/4\Delta$  for \textbf{P.3}.
\end{lemma}
\emph{Proof:} By Lemma \ref{the:constant1}, our randomized rounding technique (Algorithm \ref{alg:LPP}), taking $\mathcal{C}(x^*)$ as input, returns a grouping that achieves a profit of at least $\frac{1}{4\Delta}  \sum_{C_{ti}\in \mathcal{C}(x^*)} (x^*_{ti}\cdot \lambda_t)$, where  $\Delta$ is the size of the largest possible team in $\mathcal{C}(x^*)$. By the assumption that  $x^*$ is a $\mu$-approximate solution of \textbf{Primal LP of P.3}, we have $\sum_{C_{ti}\in \mathcal{C}(x^*)} (x^*_{ti}\cdot \lambda_t)\geq \mu\cdot OPT$. It follows that Algorithm \ref{alg:LPP1} achieves a profit of at least
\[\frac{1}{4\Delta}  \sum_{C_{ti}\in \mathcal{C}(x^*)} (x^*_{ti}\cdot \lambda_t)\geq \frac{\mu}{4\Delta}\cdot OPT.\]
 $\Box$

\begin{lemma}
\label{lem:2-1}
Assume $x^*$ is a $\mu$-approximate solution of \textbf{Primal LP of P.3} that is found by the ellipsoid method, Algorithm \ref{alg:LPP2} (whose LP rounding method is replaced with Algorithm \ref{alg:LPP}), using a modified partitioning criterion,  achieves an approximation ratio  of $\frac{\mu}{8\sqrt{f_{\max}m}}$  for \textbf{P.3}, where $f_{\max}=\max_{ u\in \mathcal{V}} f_u$.
\end{lemma}
\emph{Proof:}
To prove this lemma, it suffices to show that $\max\{\sum_{C_{ti}\in \widetilde{\mathcal{C}}} \lambda_t, \lambda_{t_{\max}}\}\geq \frac{\mu}{8\sqrt{f_{\max}m}}\cdot OPT$.

We first bound the gap between the profit gained from $\widetilde{\mathcal{C}}$  and $\sum_{C_{ti}\in  \mathcal{C}(x^*)^1}(x^*_{ti}\cdot \lambda_t)$. By Lemma \ref{the:constant2}, we have
\begin{equation}
\label{eq:1-2}\sum_{C_{ti}\in \widetilde{\mathcal{C}}} \lambda_t \geq \frac{1}{4\rho(\widetilde{\mathcal{C}})}\cdot \sum_{C_{ti} \in \mathcal{C}(x^*)^1}(x^*_{ti}\cdot \lambda_t ) \geq \frac{1}{4\sqrt{f_{\max}m}}\cdot \sum_{C_{ti}\in \mathcal{C}(x^*)^1}(x^*_{ti}\cdot \lambda_t ) \end{equation}
 where the second inequality is due to the assumption that $\rho(\widetilde{\mathcal{C}}) \leq \sqrt{f_{\max}m}$ holds for all teams from $\mathcal{C}(x^*)^1$.

We next bound the gap between the profit gained from $C_{t_{\max}}$ and $\sum_{C_{ti}\in  \mathcal{C}(x^*)^2}(x^*_{ti}\cdot \lambda_t)$. In particular, we show that
\begin{eqnarray}
\label{eq:211-2}
\lambda_{t_{\max}}\geq \frac{1}{\sqrt{m}} \cdot \sum_{C_{ti} \in \mathcal{C}(x^*)^2}(x^*_{ti}\cdot \lambda_t)
\end{eqnarray}

The following chain proves this inequality:
\begin{equation}
\label{eq:2-2}\sum_{C_{ti} \in \mathcal{C}(x^*)^2}(x^*_{ti}\cdot \lambda_t) \leq \sum_{C_{ti} \in \mathcal{C}(x^*)^2}(x^*_{ti}\cdot \lambda_{t_{\max}})=\lambda_{t_{\max}}\cdot \sum_{C_{ti} \in \mathcal{C}(x^*)^2}x^*_{ti}\leq \lambda_{t_{\max}}\cdot \sqrt{f_{\max} m}.
\end{equation}
The first inequality is due to the assumption that $C_{t_{\max}}$ delivers the highest profit among $\mathcal{C}(x^*)^2$. We next focus on proving the second inequality. Because  $x^*$ is a feasible solution of \textbf{Primal LP of P.3} and $\mathcal{C}(x^*)^2\subseteq \mathcal{C}$, we have $\sum_{C_{ti}\in \mathcal{C}(x^*)^2:  C_{ti} \ni u} x^*_{ti} \leq f_u, \forall u\in \mathcal{V}$. It follows that
\begin{eqnarray}
\label{eq:rain1}
\sum_{C_{ti} \in \mathcal{C}(x^*)^2}(x^*_{ti}\cdot |C_{ti}|)\leq \sum_{u\in \mathcal{V}} f_u  \leq f_{\max} m.
 \end{eqnarray}
Meanwhile, recall that all teams in $\mathcal{C}(x^*)^2$ contain at least $\sqrt{f_{\max}m}$ individuals, we have $\sum_{C_{ti} \in \mathcal{C}(x^*)^2}(x^*_{ti}\cdot |C_{ti}|) \geq \sum_{C_{ti} \in \mathcal{C}(x^*)^2}(x^*_{ti}\cdot \sqrt{f_{\max}m})$. This, together with (\ref{eq:rain1}), implies that
$\sum_{C_{ti} \in \mathcal{C}(x^*)^2}(x^*_{ti}\cdot \sqrt{f_{\max}m}) \leq f_{\max}m$, thus $\sum_{C_{ti}\in \mathcal{C}(x^*)^2}(x^*_{ti}) \leq \sqrt{f_{\max}m}$. This finishes the proof of the second inequality.

By the assumption that $x^*$ is a $\mu$-approximate solution of \textbf{Primal LP of P.3}, we have
\begin{equation}
\label{eq:3-1}\sum_{C_{ti} \in \mathcal{C}(x^*)}(x^*_{ti}\cdot \lambda_t) \geq \mu \cdot OPT \Rightarrow \sum_{C_{ti} \in \mathcal{C}(x^*)^1}(x^*_{ti} \cdot \lambda_t) + \sum_{C_{ti} \in \mathcal{C}(x^*)^2}(x^*_{ti} \cdot \lambda_t) \geq \mu \cdot OPT.\end{equation}

Now we are ready to prove this theorem.
\begin{eqnarray}
&&\max\{\sum_{C_{ti}\in \widetilde{\mathcal{C}}} \lambda_t, \lambda_{t_{\max}}\}\geq \frac{\sum_{C_{ti}\in \widetilde{\mathcal{C}}} \lambda_t+ \lambda_{t_{\max}}}{2}\\
&&\geq  \frac{ \frac{1}{4\sqrt{f_{\max}m}}\cdot  \sum_{C_{ti} \in \mathcal{C}(x^*)^1}(x^*_{ti} \cdot \lambda_t) + \frac{1}{\sqrt{f_{\max}m}} \cdot \sum_{C_{ti} \in \mathcal{C}(x^*)^2}(x^*_{ti}\cdot \lambda_t) }{2}\\
&&\geq\frac{1}{4\sqrt{f_{\max}m}}\cdot \frac{ \sum_{C_{ti} \in \mathcal{C}(x^*)^1}(x^*_{ti} \cdot \lambda_t) + \sum_{C_{ti} \in \mathcal{C}(x^*)^2}(x^*_{ti}\cdot \lambda_t) }{2}\\
&&\geq \frac{\mu}{8\sqrt{f_{\max}m}}\cdot OPT
\end{eqnarray}
where the second inequality is due to (\ref{eq:1-2}) and (\ref{eq:211-2}), and the last inequality is due to  (\ref{eq:3-1}). $\Box$

Lemma \ref{lem:1-1} and Lemma \ref{lem:2-1} together imply  the following theorem.
\begin{theorem}
The modified \textsf{Approx-TG} achieves an  approximation ratio of $\max\{\mu/(4\Delta), \mu/(8\sqrt{f_{\max}m})\}$ for \textbf{P.3}, where $f_{\max}=\max_{ u\in \mathcal{V}} f_u$.
\end{theorem}


\section{Conclusion}
\label{sec:conclu}
In this paper, we study the profit-driven team grouping problem. We assume a collection of tasks $\mathcal{T}$, where each task requires a specific set of skills, and yields a different profit upon completion. Individuals may collaborate with each other in the form of \emph{teams} to accomplish a set of tasks. We aim to group individuals into different teams, and assign them to different tasks, such that the total profit of the tasks that can be performed is maximized. We consider three constraints when perform grouping, and present a LP-based approximation algorithm to tackle it. We also study several extensions of this problem. Although this paper studies team grouping problem, our results are general enough to tackle a broad range of generalized cover decomposition problems.

\bibliographystyle{ijocv081}
\bibliography{reference}




\end{document}